\documentclass[twocolumn]{aastex631}

\newcommand\aastex{AAS\TeX}

\usepackage{savesym}
\savesymbol{tablenum}
\usepackage{siunitx}
\restoresymbol{SIX}{tablenum}
\usepackage{enumitem}

\newcommand{\oii}{\mbox{[\ion{O}{2}]}}
\newcommand{\oiii}{\mbox{[\ion{O}{3}]}}
\newcommand{\hei}{\mbox{\ion{He}{1}}}

\newcommand{\sii}{\mbox{[\ion{S}{2}]}}
\newcommand{\siii}{\mbox{[\ion{S}{3}]}}
\newcommand{\feii}{\mbox{[\ion{Fe}{2}]}}
\newcommand{\hb}{\mbox{H$\beta$}}
\newcommand{\ha}{\mbox{H$\alpha$}}
\newcommand{\bra}{\mbox{Br$\alpha$}}
\newcommand{\brb}{\mbox{Br$\beta$}}
\newcommand{\brg}{\mbox{Br$\gamma$}}
\newcommand{\paa}{\mbox{Pa$\alpha$}}
\newcommand{\pab}{\mbox{Pa$\beta$}}
\newcommand{\pag}{\mbox{Pa$\gamma$}}
\newcommand{\pad}{\mbox{Pa$\delta$}}
\newcommand{\lya}{\mbox{Ly$\alpha$}}

\newcommand\msun{\mbox{\si{M_\odot}}}

\newcommand\smpy{\mbox{\si{M_\odot.yr^{-1}}}}
\newcommand\mstar{\mbox{$M_\mathrm{star}$}}

\renewcommand{\micron}{\si{\micro\meter}}
\newcommand{\zsp}{\mbox{$z_\mathrm{spec}$}}
\newcommand{\zph}{\mbox{$z_\mathrm{phot}$}}

\newcommand{\sap}{\mbox{\texttt{SAPPHIRES}}}
\newcommand{\macs}{\mbox{MACS\,J0416.1--2403}}

\newcommand{\textblue}[1]{{#1}}

\begin{document}

\title{Slitless Areal Pure-Parallel HIgh-Redshift Emission Survey (SAPPHIRES):\\
Early Data Release of Deep JWST/NIRCam Images and Spectra in MACS\,J0416 Parallel Field
}

\correspondingauthor{Fengwu Sun}
\email{fengwu.sun@cfa.harvard.edu}

% \suppressAffiliations

\author[0000-0002-4622-6617]{Fengwu Sun}
\affiliation{Center for Astrophysics $|$ Harvard \& Smithsonian, 60 Garden St., Cambridge, MA 02138, USA}

\author[0000-0001-7440-8832]{Yoshinobu Fudamoto}
\affiliation{Center for Frontier Science, Chiba University, 1-33 Yayoi-cho, Inage-ku, Chiba 263-8522, Japan}
\affiliation{Steward Observatory, University of Arizona, 933 N Cherry Avenue, Tucson, AZ 85721, USA}

\author[0000-0001-6052-4234]{Xiaojing Lin}
\affiliation{Department of Astronomy, Tsinghua University, Beijing 100084, China}
\affiliation{Steward Observatory, University of Arizona, 933 N Cherry Avenue, Tucson, AZ 85721, USA}

\author[0000-0003-4337-6211]{Jakob M. Helton}
\affiliation{Steward Observatory, University of Arizona, 933 N Cherry Avenue, Tucson, AZ 85721, USA}

\author[0000-0003-4512-8705]{Tiger Yu-Yang Hsiao}
\affiliation{Center for Astrophysics $|$ Harvard \& Smithsonian, 60 Garden St., Cambridge, MA 02138, USA}
\affiliation{Center for Astrophysical Sciences, Department of Physics and Astronomy, The Johns Hopkins University, 3400 N Charles St. Baltimore, MD 21218, USA}
\affiliation{Space Telescope Science Institute (STScI), 3700 San Martin Drive, Baltimore, MD 21218, USA}

\author[0000-0003-1344-9475]{Eiichi Egami}
\affiliation{Steward Observatory, University of Arizona, 933 N Cherry Avenue, Tucson, AZ 85721, USA}

\author[0009-0003-7532-3197]{Arshia Akhtarkavan}
\affiliation{Steward Observatory, University of Arizona, 933 N Cherry Avenue, Tucson, AZ 85721, USA}

\author[0000-0002-8651-9879]{Andrew J.\ Bunker}
\affiliation{Department of Physics, University of Oxford, Denys Wilkinson Building, Keble Road, Oxford OX1 3RH, U.K.}

\author[0000-0001-8467-6478]{Zheng Cai}
\affiliation{Department of Astronomy, Tsinghua University, Beijing 100084, China}

\author[0000-0002-4781-9078]{Christa DeCoursey}
\affiliation{Steward Observatory, University of Arizona, 933 N Cherry Avenue, Tucson, AZ 85721, USA}

\author[0000-0002-2929-3121]{Daniel J.\ Eisenstein}
\affiliation{Center for Astrophysics $|$ Harvard \& Smithsonian, 60 Garden St., Cambridge, MA 02138, USA}

\author[0000-0003-3310-0131]{Xiaohui Fan}
\affiliation{Steward Observatory, University of Arizona, 933 N Cherry Avenue, Tucson, AZ 85721, USA}

\author[0000-0002-6047-430X]{Yuichi Harikane}
\affiliation{Institute for Cosmic Ray Research, The University of Tokyo, 5-1-5 Kashiwanoha, Kashiwa, Chiba 277-8582, Japan}

\author[0000-0001-7673-2257]{Zhiyuan Ji}
\affiliation{Steward Observatory, University of Arizona, 933 N Cherry Avenue, Tucson, AZ 85721, USA}

\author[0000-0002-5768-738X]{Xiangyu Jin}
\affiliation{Steward Observatory, University of Arizona, 933 N Cherry Avenue, Tucson, AZ 85721, USA}

\author[0000-0003-3762-7344]{Weizhe Liu} % (刘伟哲)}
\affiliation{Steward Observatory, University of Arizona, 933 N Cherry Avenue, Tucson, AZ 85721, USA}

\author[0000-0003-4247-0169]{Yichen Liu} 
\affiliation{Steward Observatory, University of Arizona, 933 N Cherry Avenue, Tucson, AZ 85721, USA}

\author[0009-0003-5402-4809]{Zheng Ma}
\affiliation{Steward Observatory, University of Arizona, 933 N Cherry Avenue, Tucson, AZ 85721, USA}

\author[0000-0002-4985-3819]{Roberto Maiolino}
\affiliation{Kavli Institute for Cosmology, University of Cambridge, Madingley
Road, Cambridge CB3 0HA, UK}
\affiliation{Cavendish Laboratory, University of Cambridge, 19 JJ Thomson
Avenue, Cambridge CB3 0HE, UK}
\affiliation{Department of Physics and Astronomy, University College London,
Gower Street, London WC1E 6BT, UK}

\author[0000-0002-1049-6658]{Masami Ouchi}
\affiliation{National Astronomical Observatory of Japan, 2-21-1 Osawa, Mitaka, Tokyo 181-8588, Japan}
\affiliation{Institute for Cosmic Ray Research, The University of Tokyo, 5-1-5 Kashiwanoha, Kashiwa, Chiba 277-8582, Japan}
\affiliation{Department of Astronomical Science, SOKENDAI (The Graduate University for Advanced Studies), Osawa 2-21-1, Mitaka, Tokyo, 181-8588, Japan}
\affiliation{Kavli Institute for the Physics and Mathematics of the Universe (WPI), University of Tokyo, Kashiwa, Chiba 277-8583, Japan}

\author[0000-0003-0747-1780]{Wei Leong Tee}
\affiliation{Steward Observatory, University of Arizona, 933 N Cherry Avenue, Tucson, AZ 85721, USA}

\author[0000-0002-7633-431X]{Feige Wang}
\affiliation{Department of Astronomy, University of Michigan, 1085 S. University Ave., Ann Arbor, MI 48109, USA}

\author[0000-0001-9262-9997]{Christopher N. A. Willmer}
\affiliation{Steward Observatory, University of Arizona, 933 N Cherry Avenue, Tucson, AZ 85721, USA}

\author[0000-0003-0111-8249]{Yunjing Wu}
\affiliation{Department of Astronomy, Tsinghua University, Beijing 100084, China}

\author[0000-0002-5768-8235]{Yi Xu}
\affiliation{Institute for Cosmic Ray Research, The University of Tokyo, 5-1-5 Kashiwanoha, Kashiwa, Chiba 277-8582, Japan}
\affiliation{Department of Astronomy, Graduate School of Science, the University of Tokyo, 7-3-1 Hongo, Bunkyo, Tokyo 113-0033, Japan}

\author[0000-0001-5287-4242]{Jinyi Yang}
\affiliation{Department of Astronomy, University of Michigan, 1085 S. University Ave., Ann Arbor, MI 48109, USA}

\author[0000-0002-1574-2045]{Junyu Zhang}
\affiliation{Steward Observatory, University of Arizona, 933 N Cherry Avenue, Tucson, AZ 85721, USA}

\author[0000-0003-3307-7525]{Yongda Zhu}
\affiliation{Steward Observatory, University of Arizona, 933 N Cherry Avenue, Tucson, AZ 85721, USA}

% \author{SAPPHIRES Team (please add your info on overleaf \texttt{00\_main.tex})}

\begin{abstract}

We present the early data release (EDR) of \sap, a JWST Cycle-3 Treasury imaging and spectroscopic survey using the powerful NIRCam wide-field slitless spectroscopic (WFSS) mode in pure parallel.
\sap\ will obtain NIRCam imaging and WFSS data in many cosmological deep fields totaling a telescope charged time of 709 hours (557-hour exposures).
In this EDR, we present NIRCam imaging and WFSS data obtained in parallel to the Frontier Field galaxy cluster MACS\,J0416.1--2403, which are attached to primary observations JWST-GO-4750.
With a total dual-channel exposure time of 47.2 hours, we obtain deep NIRCam imaging in 13 bands at 0.6--5.0\,\micron\ and deep WFSS at 3.1--5.0\,\micron\ through the F356W and F444W filters with grisms in orthogonal dispersion directions.
We release reduced NIRCam images, photometric catalogs of 22107 sources and WFSS spectra of \textblue{1060} sources with confirmed redshifts ($z\simeq0-8.5$).
Preliminary value-added catalogs including photometric redshifts, spectroscopic redshifts and physical properties (mass, star-formation rate, etc.) are also made available.
We also characterize the data quality and demonstrate scientific applications, including (1) galaxy candidates at the redshift frontier ($z\gtrsim10$), (2) the ionized gas kinematics of a galaxy reconstructed from $R\sim1500$ grism spectra at orthogonal dispersion directions, (3) massive emission-line galaxies and active galactic nuclei (AGN) around the Epoch of Reionization.
\end{abstract}


\section{Introduction}
\label{sec:01_intro}

The long-wavelength (LW; 2.4--5.0\,\micron) grisms on board of JWST's primary imager, the near-infrared camera \citep[NIRCam;][]{rieke23a} were initially developed for telescope primary segment phasing and  wavefront sensing \citep{greene17}.
Although designed as an engineering observing mode, NIRCam LW wide-field slitless spectroscopy (WFSS) has been approved for primary observations since Cycle 1 (2022).
NIRCam WFSS has a lower sensitivity compared with JWST's primary near-infrared spectrograph \citep[NIRSpec;][]{jakobsen22} which uses slits to reduce the sky background.
However, NIRCam WFSS can effectively obtain spectra of all galaxies entering its wavelength-dependent field of view (FoV). %  which is ideal for flux-complete selection of emission-line galaxies at high redshifts.
Through the commissioning phase of JWST, it has been revealed that the systematic throughput of NIRCam WFSS at 3--5\,\micron\ is higher than the pre-launch estimate up to $\sim40$\% \citep{rieke23a,rigby23}.
This enables the detections of strong emission-line galaxies at $z>6$ through \ha\ and \oiii\,$\lambda$5008 lines even with shallow (10--20\,min) flux-calibration data \citep{sun22b,sun23}.

Starting with JWST Cycle-1, NIRCam WFSS has been offering a huge amount of exciting scientific discoveries.
This includes the unprecedented characterization of ices in molecular cloud complexes within the Milky Way \citep[e.g.,][]{mcclure23,noble24,Smith_IceAge}.
Beyond the local Universe, NIRCam WFSS has been discovering emission-line galaxies, broad-line active galactic nuclei (AGNs), dusty star-forming galaxies and the large-scale structure that they trace up to $z\sim9$ \citep[e.g.,][]{sun22b,sun23,sun24,sun25a,kashino23,matthee23,matthee24,wangf23,champagne24,helton24a,helton24b,hd24,linx24,xiao24}.
Cycle-1 NIRCam WFSS surveys (e.g., ASPIRE, \citealt{wangf23,yangj23}; CEERS, \citealt{finkelstein25}; EIGER, \citealt{kashino23}; FRESCO, \citealt{oesch23}) have successfully delivered the first look at the Epoch of Reionization (EoR; $z\gtrsim 6$) traced by rest-frame optical nebular emission lines.
Standing on the shoulders of these giants, several Cycle-2 NIRCam WFSS programs focus on gravitationally lensed galaxies behind massive lensing clusters (e.g., ALT, \citealt{naidu24}; MAGNIF, Sun, F.\ et al.\ in prep.; GO-3538, PI: Iani, E.; GO-4043, PI: Witten, C.).
These observations are delivering the deep insight of the dwarf galaxy population at the EoR, which is thought to play a major role in the cosmic reionization (e.g., see a review by \citealt{robertson22}).

Existing Cycle-1/2 NIRCam WFSS surveys have been typically limited by the survey area ($\sim 100$\,arcmin$^2$).
The first two years of JWST observations suggest that there are significant populations of luminous (massive) galaxies and AGNs at high redshifts, including $z\gtrsim10$ Lyman-break galaxies \citep[e.g.,][]{bunker23,finkelstein23,harikane23a,hainline24,carniani24a}, $z\gtrsim4$ quiescent galaxies, \citep[e.g.,][]{carnall23b,valentino23,long24}; $z\gtrsim5$ massive star-forming galaxies, \citep[e.g.,][]{gottumukkala24,weibel24a,xiao24}, and a large population of AGNs beyond the so-called ``Cosmic Noon'' \citep[$z\gtrsim 3$; e.g.,][]{harikane23b,maiolino24a,matthee24}.
Wide-field imaging and spectroscopic surveys have been recognized as one of the best approaches to characterize the abundance of these systems and enable further detailed follow-up studies.
In JWST Cycle-3, there are four large treasury general observer (GO) programs that aim to conduct wide-field near-IR imaging and spectroscopic surveys of the high-redshift Universe.
One of these large treasury programs is CAPERS (GO-6368; PI: Dickinson, M.), which will perform NIRSpec multi-object spectroscopy (MOS) with prism ($\lambda = 0.6-5.3$\,\micron) as the primary observing mode. The other three large treasury programs are COSMOS-3D (GO-5893; PI: Kakiichi, K.), NEXUS (GO-5105; PI: Shen, Y.; \citealt{sheny24}) and \sap\ (GO-6434; PI: Egami, E.), all of which will obtain WFSS with NIRCam across large areas and variety of sightlines.

The Slitless Areal Pure-Parallel HIgh-Redshift Emission Survey (\sap) is the first JWST Treasury GO program that uses NIRCam WFSS in the pure parallel mode, which has become available since Cycle 3.
Together with GO-5398 POPPIES (PI: Kartaltepe, J.\ \& Rafelski, M.), NIRCam WFSS pure-parallel observations in Cycle 3 will deliver the essential imaging and spectroscopic surveys over a large volume and independent sightlines (i.e., mitigating shot noise and cosmic variance) at no cost to the primary observing time of JWST.
\sap\ was designed to conduct a highly complete (in terms of emission-line fluxes) spectroscopic survey of the most luminous ($M_\mathrm{UV} \lesssim -21$\,AB mag) galaxies at $z=4-9$ and potentially beyond, thereby making a robust determination of their number density and their role in galaxy evolution around the EoR.
As a treasury program, \sap\ will also study the population of broad-line AGN through \ha, \hb\ and Paschen / \hei\ lines at lower redshifts, capture galaxies overdensities and underdensities across $z\simeq 1 - 9$, and offer spatially resolved views of the emission-lines and kinematics of galaxies through medium-resolution grism observations at $R\simeq1300-1700$.

In this paper, we present the early data release of \sap, which was obtained as NIRCam imaging and WFSS parallel to GO-4750 (PI: Nakajima, K.), NIRSpec MOS observations in the \macs\ field.
\macs\ is one of the six Hubble Frontier Field clusters \citep{lotz17}.
We describe the overall design and execution of \sap\ observations in Section~\ref{sec:02_obs}.
NIRCam imaging and WFSS data processing techniques are described in Section~\ref{sec:03_ana}.
We present the construction of value-added data products (VAD) in Section~\ref{sec:04_vad}, and a few demonstrations of scientific applications in Section~\ref{sec:05_sci}.
The summary of the released data and potential caveats are presented in Section~\ref{sec:06_data}.
The paper is concluded in Section~\ref{sec:07_sum}.
Throughout this work, we assume a flat $\Lambda$CDM cosmology with $H_0= 70$\,\si{km.s^{-1}.Mpc^{-1}} and $\Omega_\mathrm{M} = 0.3$.

\section{Observations}
\label{sec:02_obs}

\subsection{\sap\ Program Design}
\label{ss:2a_design}

\sap\ (GO-6434; PI: Egami, E.; co-PIs: Fan, X., Sun, F., Wang, F.\ and Yang, J.) is a JWST Cycle-3 Treasury large program that obtains NIRCam grism spectroscopy in the pure parallel mode.
When NIRCam obtains grism spectra in the LW channel, the short-wavelength (SW; 0.6--2.3\,\micron) channel is used to simultaneously obtain imaging observations at the same pointing, and the data will be used to calibrate the astrometry of LW spectra and interpret the spectral energy distributions (SEDs) of the sources selected with the LW grism. 
Therefore, \sap\ is also a pure-parallel NIRCam imaging survey over a wide survey area and many independent sightlines.

Given the pure-parallel nature of our program, \sap\ cannot determine the exact pointings of our surveys, which are fully dependent on the pointings and roll angles of primary observations. 
The dither patterns, exposure time and number of filters / grisms of \sap\ observations also depend on those of the primary observations.
For example, if the primary observation is NIRSpec spectroscopy with standard three-point dithering, no change of filter / grating and no split of visit / configuration, our pure-parallel observation attachment will be a \textit{``monolithic''} NIRCam WFSS observation, including one-filter LW WFSS and one-filter SW imaging at just three dithering positions.
Without LW imaging, it will be difficult to fully photometrically select the galaxies that are responsible for continuum and line emissions detectable within the grism data.
However, a significant number of \sap\ observations will fall on top of existing JWST NIRCam$+$NIRISS imaging footprints in well-studied cosmological deep fields, and we are able to interpret these monolithic grism data with existing images.

As of December 2024, the \sap\ team has designed the parallel observation according to the primary observing programs that the team is allowed to attach to.
\sap\ observations are planned in all five CANDELS fields (\citealt{grogin11,koekemoer12}, including GOODS-S, GOODS-N, EGS, UDS and COSMOS) and Abell 2744, where extensive HST and JWST imaging data coverage have been collected over the past decades \citep[e.g.,][]{giavalisco04,davis07,lawrence07,scoville07,postman12,lotz17,treu22,casey23,eisenstein23,rieke23b,oesch23,bezanson24,suess24,naidu24,finkelstein25}.
The total charged time of \sap\ is 709\,hours (excluding skipped observations; 557-hour exposures), including {375} hours for WFSS exposures and {182} hours for pure dual-channel imaging exposures.
The median WFSS exposure time for all planned observations is 2.1\,hour, and 18\% of the WFSS observations have an exposure time longer than 5 hours (overlapped footprints are not considered).

With the depths and the scope of our planned observations, we prioritize WFSS observations in the F444W filter in our design, and we opt to use the row-direction grism (Grism R) for simpler 2D background subtraction and less spectral confusion from sources outside of the detector FoV. 
When the \sap\ WFSS footprint is expected to overlap with existing or planned F444W WFSS observations at similar depth (e.g., GO-5893 COSMOS-3D observations), we prioritize the F356W filter instead.
If more mechanism moves are available for the same primary observations, we conduct (1) multiple-wide-band NIRCam imaging, (2) F356W$+$F444W Grism R spectroscopy, (3) multiple-medium-band NIRCam imaging and (4) Grism C spectroscopy with a descending order of priorities.
The multiple-band NIRCam imaging is to better determine the SEDs of grism sources, the F356W$+$F444W spectroscopy is to improve the wavelength coverage (and thus survey volume in redshift space), and the use of Grism R$+$C will improve the confusion mitigation and allow further kinematic studies.
We opt not to use the F277W filter for WFSS observations because the second-order spectra therein are not well calibrated in Cycle-3.
For primary observations with interesting targets at specific redshifts (e.g., the SSA22 field that contains a well-known galaxy protocluster at $z=3.1$, \citealt{steidel98,umehata15,umehata19}), we change our observation strategy accordingly (e.g., F322W2 filter for WFSS observations in SSA22 field because \ha\ at $z=3.1$ falls at 2.7\,\micron).

The detailed observation designs including target coordinates, filter / pupil usage, exposure times and dither patterns will be presented in a future paper from the \sap\ collaboration once the data acquisition is complete.
On the public page of our program information\footnote{\url{https://www.stsci.edu/jwst-program-info/download/jwst/pdf/6434/}}, we have grouped our planned observations into folders that are named after the primary program ID, field of observation and primary program PI name if the observation falls in cosmological deep fields.
Taking the EDR observation as an example, our observations are grouped into folders \textsf{``4750-M0416-Nakajima-1/2/3''} (Primary Program ID: 4750, PI: Nakajima, K.\ in M0416 field).
The names of certain observation folders start with prefixes.
Monolithic primary observations are indicated by a prefix of \textsf{``x''} in the observation folder name, and high-data rate primary observations ($>$1\,MB\,s$^{-1}$) are indicated by a prefix of \textsf{``d''}.

The observation labels of \sap\ are named after the primary program ID, observation number, visit number, exposure sequence and type of parallel observation.
For example, our observation number 111 ``\textsf{4750003001a-SP}'' is attached to GO-4750, observation number 003, visit number 001 and first exposure, which is a WFSS exposure (\textsf{SP} for spectroscopy, and \textsf{IM} for imaging).

% Figure~\ref{fig:footprint_all} shows the predicted \sap\ footprints in five CANDELS fields (REF\#, including GOODS-S, GOODS-N, EGS, UDS and COSMOS) and Abell 2744.
% These well-studied fields have extensive HST and JWST imaging data coverage over decades (REF\#).

% \begin{figure*}[!t]
% \centering
% \includegraphics[width=\linewidth]{figures/sapphires_footprint_202411.pdf}
% \caption{Preliminary SAPPHIRES footprints in cosmological deep fields as of December 2024.}
% \label{fig:footprint_all}
% \end{figure*}

\subsection{EDR Observations}
\label{ss:2b_4750}

The EDR of \sap\ makes use of the pure-parallel NIRCam imaging and spectroscopic observations attached to JWST Cycle-3 GO-4750 (PI: Nakajima, K.), which obtained NIRSpec MOS observations in the \macs\ field over November 4--6, 2024.
With a total charge time of 63.2 hours, GO-4750 was split into three visits, and a total of 20 mechanism moves are available from the primary observations.
GO-4750 used the standard three-shutter nodding with NIRSpec, each with an integration time of 2947\,seconds.
Given the primary observation design, we obtained NIRCam observation with all 20 mechanism moves, each with five-group three-integration \textsc{deep8} readout (effective exposure time is 2834\,seconds per dither position).
The only exception is our observation number 134, where the NIRCam F140M$+$F335M imaging was obtained with seven-group two-integration \textsc{deep8} readout to mitigate a data rate excess issue (effective exposure time is 2748\,seconds per dither position).
Therefore, the exposure time associated with each mechanism move is 8.5 or 8.2\,ksec.

\sap\ EDR NIRCam imaging observations were obtained with 13 filters, including F070W, F090W, F115W, F140M, F150W, F182M, F200W, F210M (SW), F277W, F335M, F356W, F410M and F444W (LW) over a total area of 16.9\,arcmin$^2$.
Table~\ref{tab:01_depth} presents the detailed area and total exposure time of the imaging observation with each filter.
SW imaging data were obtained for all twenty mechanism moves.
% These are not only used for constructing multiple-wavelength SEDs of detected sources, but also for quantifying the astrometric errors in the LW grism data, which are found to be small because the primary NIRSpec observations.
Among all mechanism moves, ten of them are spent for LW imaging and the remaining ten are for LW WFSS.

The ten LW WFSS mechanism moves consist of three for F444W Grism R, three for F444W Grism C, two for F356W Grism R and two for F356W Grism C.
The total WFSS exposure time is 7.1 hours for F444W in each dispersion direction and 4.7 hours for F356W in each dispersion direction (Table~\ref{tab:02_grism}).
This makes \sap\ EDR the deepest NIRCam WFSS survey by Cycle 3 with complete 3.1--5.0\,\micron\ spectral coverage at two dispersion directions.

\begin{figure*}[!th]
\centering
\includegraphics[width=\linewidth]{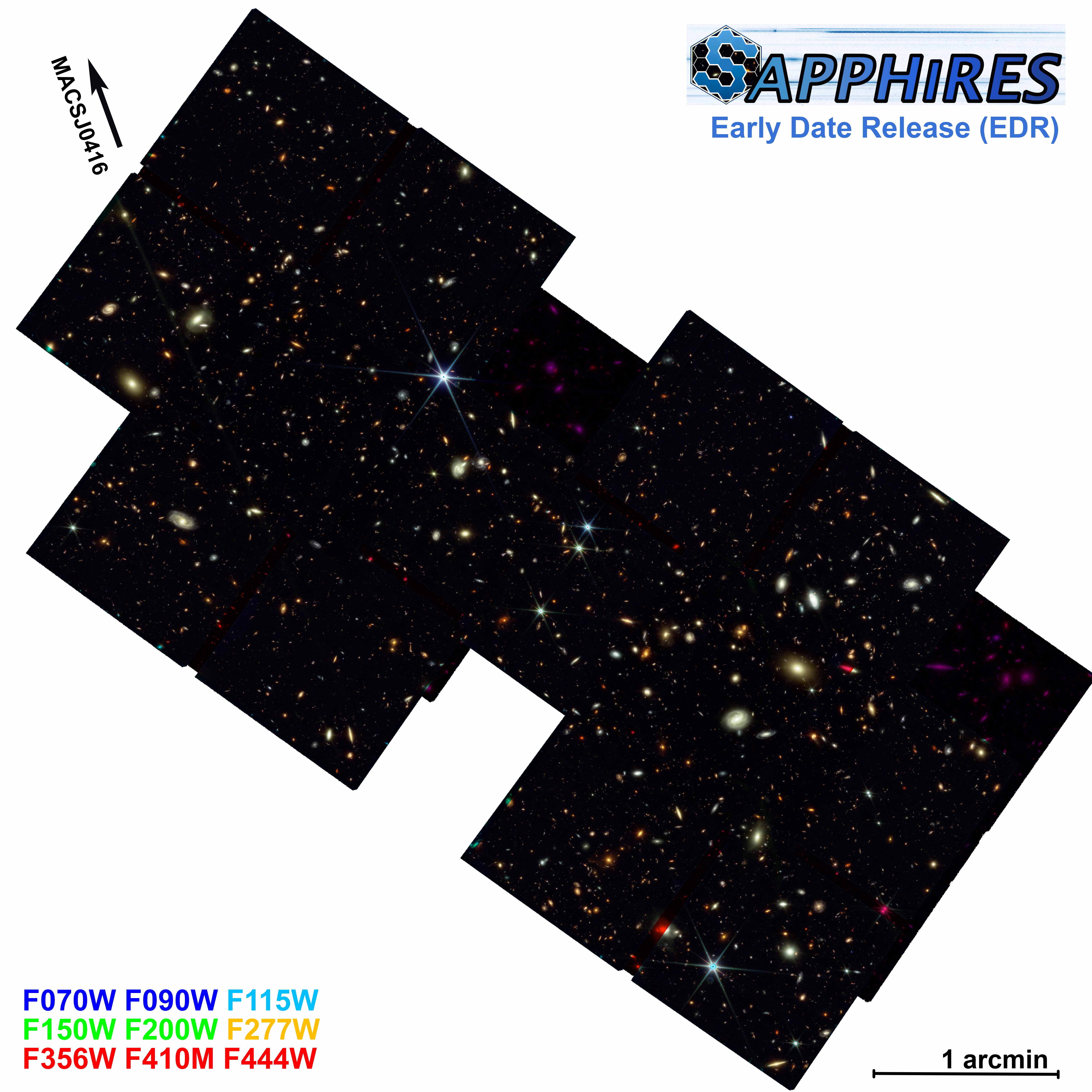}
\caption{JWST/NIRCam false-color image of the \sap\ EDR mosaics in the MACS\,J0416 parallel field. This image is made using nine-band NIRCam data and the color scheme is shown in the bottom-left corner. North is up, east is left. The direction of MACS\,J0416 cluster center is indicated in the top-left corner, and the scale is shown in the bottom-right corner. Certain regions of the mosaics are not fully covered with these bands because of detector gaps and imperfect mosaic patterns.
}
\label{fig:mosaics}
\end{figure*}

We also examine the footprint of \sap\ EDR observations against existing data on Mikulski Archive for Space Telescopes (MAST\footnote{\url{https://archive.stsci.edu/}}).
Although the primary target MACS\,J0416.1--2403 is a well studied Frontier Field cluster that has been observed through multiple JWST imaging-spectroscopic surveys (e.g., CANUCS, \citealt{willott22}; PEARLS, \citealt{windhorst23}; also by the pure-parallel program PANORAMIC; \citealt{panoramic}), the only JWST imaging data that overlap with our footprint is the F210M$+$F480M images taken with JWST-GO-2883 (MAGNIF; PI: Sun, F.; see also \citealt{fus25}) as shallow out-of-field imaging for the grism observation.
Given the large contrast in depth ($\gtrsim10\times$ in exposure time) and small overlap ($\sim2$\,arcmin$^2$), MAGNIF data are not included in the analyses.
We also confirm that the \sap\ footprint is outside the HST Frontier Field survey footprint \citep{lotz17}, and only has a small overlap ($\lesssim1$\,arcmin$^2$) with the wider but shallower HST BUFFALO survey \citep{steinhardt20}.
Therefore, the EDR is purely based on the JWST data taken through our program and no archival HST or JWST data are used.

\section{Data Processing}
\label{sec:03_ana}

\subsection{NIRCam Imaging Data Reduction}
\label{ss:3a_img}

% \textblue{Yoshi: Please check and modify - thanks!}

The \sap\ EDR NIRCam imaging data were processed using a customized \textsc{jwst} pipeline \citep{bushouse24} \verb|v1.11.2| with the reference files \verb|jwst_1293.pmap|. % (including JWST Cycle-1 NIRCam flux calibrations).
We first processed all NIRCam imaging and WFSS data through the standard stage-1 pipeline.
After the standard stage-2 pipeline where the images were flat-fielded and photometrically calibrated, we introduced customized steps including (\romannumeral 1) 1/f noise stripe subtraction in both row and column directions, 
(\romannumeral 2) wisp removal in NIRCam SW detectors using the STScI templates, (\romannumeral 3) bad-pixel masking using the sigma-clipped median images taken in each detector, 
(\romannumeral 4) median sky background subtraction and
(\romannumeral 5) world coordinate system (WCS) correction using DESI Legacy Imaging Survey \citep{dey19} source catalog, which has been tied to the astrometry of Gaia DR2 \citep{gaiadr2}.
Most of these customized steps are designed to correct for well-known artifacts identified since telescope commissioning \citep{rigby23}, and these steps have been widely used for NIRCam imaging data processing by successful extragalactic programs like CEERS \citep{bagley23,finkelstein23}, JADES \citep{rieke23b,eisenstein23b} and many others.
To construct the bad-pixel templates, we first median-filtered each image with 3$\times$3-pixel kernel and subtracted off the filtered image.
Bad pixels were then identified on the sigma-clipped median image taken in each detector.
To subtract the median sky background, we first masked the top 25th-percentile rows and columns sorted by decreasing median brightness, and then computed the median background from each image.
The use of DESI Legacy survey catalog mitigates the issues of limited \textit{Gaia} stars entering the NIRCam FoV.
We also processed the F444W image mosaics first to construct a source catalog to correct for the astrometric errors seen in bluer images.
The astrometric errors are found to be small ($\lesssim0\farcs02$) because careful target acquisition has been conducted for the primary NIRSpec observations.

Calibrated NIRCam images in each band were mosaicked the through standard stage-3 pipeline with an output pixel size of 0\farcs03, \texttt{pixfrac = 1.0} and a common north-up-east-left WCS frame for all filters. 
Figure~\ref{fig:mosaics} shows the false-color image constructed using nine-band \sap\ EDR mosaics.
% The quality of our NIRCam mosaics is discussed in the following subsection.

\subsection{Photometric Catalog}
\label{ss:3b_cat}

To conduct source detection for grism spectroscopic analyses, we first constructed an inverse-variance stacked image using data from eight NIRCam bands (F182M--F444W).
The stacked image was then iteratively median-filtered six times. 
This step is to suppress the extended emission for bright large galaxies and enhance the detection for faint blended sources, but at the expense of overshredding large galaxies with substructures and complex morphology.
In each iteration we (\romannumeral1) subtracted the median-filtered (box size: 0\farcs57$\times$0\farcs57) image from the stacked images, (\romannumeral2) identified the pixels in residuals that deviate from zero at $>2.5\sigma$, (\romannumeral3) made a temporary image from the detection image but replaced the strong-residual pixels with the median-filtered image, and (\romannumeral4) median-filtered the temporary image for next iteration. 
We then applied a manual mask of diffraction spikes to this sharpened stacked image.
Such a detection image was then input into a \texttt{photutils} pipeline \citep{photutils}, in which we generated the segmentation map at a S/N (per pixel) threshold of 3.0 and a minimal number of connecting pixels of 10.
The segmentation map was then deblended using \texttt{deblend\_sources} command with the parameter set \texttt{npixels=10}, \texttt{nlevels=32}, and \texttt{contrast=0.01}.
An initial source catalog was generated on the stacked image (without sharpening or spike-masking) with the deblended segmentation map.

Aperture photometry of detected sources was performed on 13-band NIRCam images using four sets of circular apertures (\texttt{circ1}: $r=0\farcs10$; \texttt{circ2}: $r=0\farcs15$; \texttt{circ3}: $r=0\farcs25$; \texttt{circ4}: $r=0\farcs30$) and two sets of Kron apertures (\texttt{kron}: Kron parameter $K=2.5$; \texttt{kron\_s}: $K=1.2$).
Local background was subtracted using a rectangular annulus with a width of 0\farcs15.
Point-source aperture loss was corrected using \texttt{webbpsf} models \citep{webbpsf}.

The photometric uncertainty was computed through two methods following the JADES practice \citep{rieke23b}, one based on the error extension of the images (with suffix \texttt{\_en} in our catalog) and one based on random aperture experiments (with suffix \texttt{\_e}).
In each imaging band, we injected random apertures at fixed radii (from 0\farcs06 to 1\arcsec) in source-free areas.
Flux densities ($f$) from the scientific extension and mean weights ($\bar{w}$) from the weight extension are measured from these apertures.
At each radius, we derived the standard deviation of random-aperture flux densities ($\sigma_f$) within ten bins of mean weights ($\bar{w}$) within the apertures. 
We fit the relation between $\sigma_f$ and $\bar{w}$ using a power law: $\sigma_f = \sigma_0 \bar{w}^{\alpha}$, where $\alpha$ is typically measured around $-0.5$ as expected.
We then fit the radius dependence of $\sigma_0$ and $\alpha$ with a power law, and then computed the flux density uncertainty for each aperture using the measured $\bar{w}$ and best-fit $\sigma_0(r)$ and $\alpha(r)$.
The flux density uncertainties based on random aperture experiments ($f_\mathrm{e}$) are typically found to be larger than those based only on error extension ($f_\mathrm{en}$) because of a more realistic treatment of correlated noise.
However, for bright sources ($\lesssim21$\,AB mag), $f_\mathrm{en}$ are generally larger than $f_\mathrm{e}$ because the error extension of mosaicked images have properly included the Poisson noise from the bright sources.
In our later analyses, we use the larger photometric uncertainty derived with either of the two methods.

\begin{figure*}[!th]
\centering
\includegraphics[width=\linewidth]{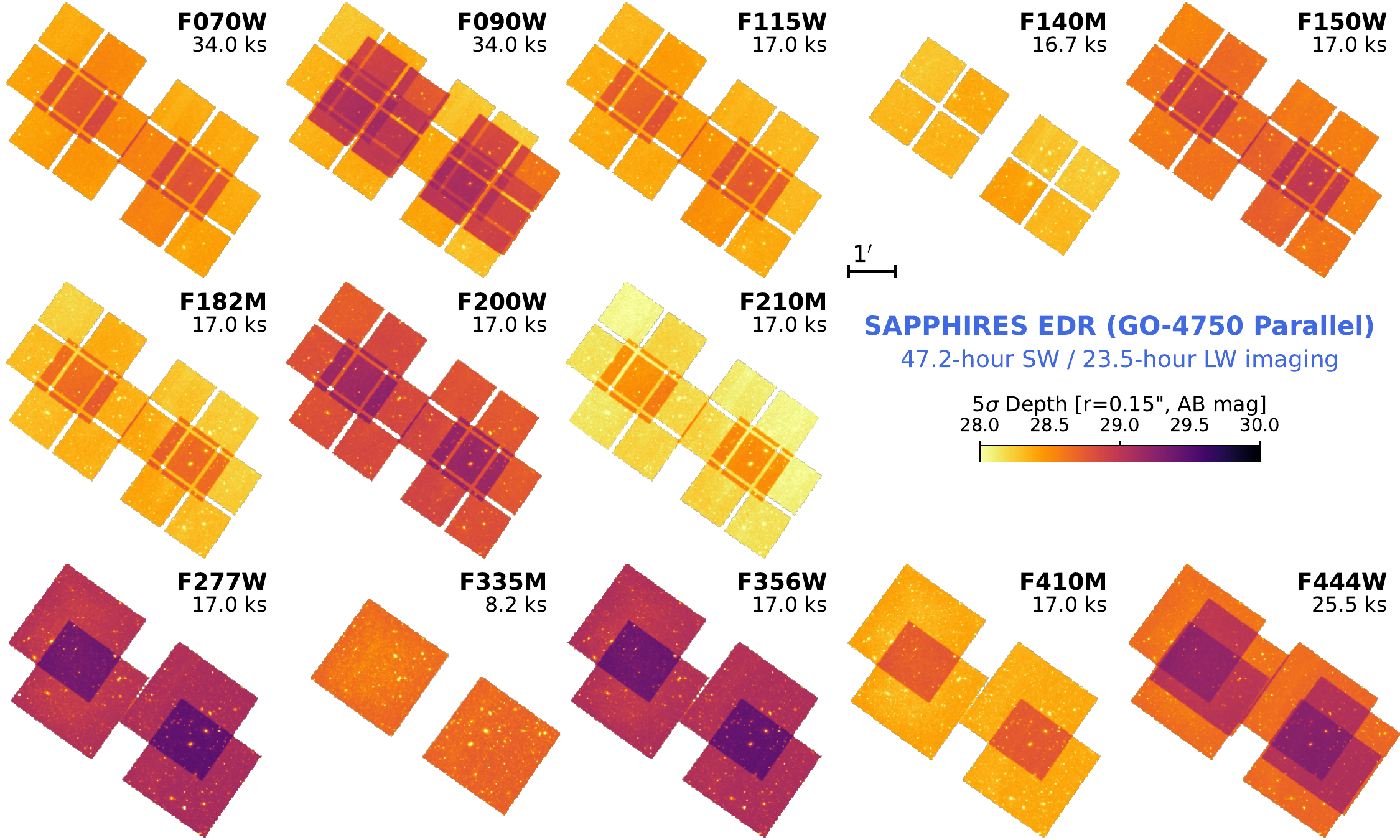}
\caption{$5\sigma$ depth (unit: AB mag) maps for point sources through aperture photometry ($r=0\farcs15$) measured with 13-band \sap\ EDR observations.
These maps are computed from the error extension of the imaging mosaics (scaled to the depths measured with random aperture experiments) with aperture correction.
The filter name and total exposure time are shown in the top-right corner of each map.
}
\label{fig:depth_map}
\end{figure*}

Table~\ref{tab:01_depth} summarizes the 5$\sigma$ depth of point sources measured with $r=0\farcs15$ aperture (i.e., \texttt{circ2}), computed using the random-aperture method. 
The $5\sigma$ depths are also displayed in Figure~\ref{fig:depth_map}.
The \sap\ EDR data reach $5\sigma$ depths of $\sim29.0$\,AB mag in most of the wide bands.
With \texttt{circ2} aperture, the greatest depth is seen in the F356W band, reaching $\sim$29.8\,AB mag in the deepest region.

To reject hot pixels or cosmic rays that still persist in our detection image and thus catalog, we require $\geq5\sigma$ detections in at least two NIRCam bands. 
This results in a total of 22107 entries in our imaging source catalog.

\begin{table}[!t]
\centering
\begin{tabular}{ccccc}
\hline\hline
Filter & $t_\mathrm{exp}$ [s] & Area [arcmin$^2$] & $5\sigma$ Depth [AB mag]  \\
 % & [s] & [arcmin$^2$] & [AB mag] \\
(1) & (2) & (3) & (4) \\
\hline
F070W & 34014 & 15.3 & 28.6--28.7--29.0 \\
F090W & 34014 & 16.7 & 28.5--28.9--29.3 \\
F115W & 17007 & 15.3 & 28.5--28.6--29.0 \\
F140M & 16749 &  9.1 & 28.5--28.5--28.6 \\
F150W & 17007 & 15.3 & 28.8--28.8--29.2 \\
F182M & 17007 & 15.3 & 28.4--28.5--28.9 \\
F200W & 17007 & 15.3 & 28.9--29.0--29.4 \\
F210M & 17007 & 15.3 & 28.3--28.4--28.7 \\
% \hline
F277W & 17007 & 15.3 & 29.3--29.4--29.7 \\
F335M &  8245 &  9.3 & 28.9--29.1--29.1 \\
F356W & 17007 & 15.3 & 29.4--29.5--29.8 \\
F410M & 17007 & 15.3 & 28.7--28.8--29.2 \\
F444W & 25510 & 16.5 & 29.0--29.1--29.6 \\
\hline
\end{tabular}
\caption{Summary of \sap\ EDR imaging mosaics properties. Column (1): NIRCam filter; (2): total exposure time in unit of seconds; (3): total area in unit of arcmin$^2$; (4): $5\sigma$ depth of point source measured with $r=0\farcs15$ aperture at 10--50--90th percentiles of the area in each band (see Section~\ref{ss:3a_img} for details). }
\label{tab:01_depth}
\end{table}

\begin{figure*}[!th]
\centering
\includegraphics[width=\linewidth]{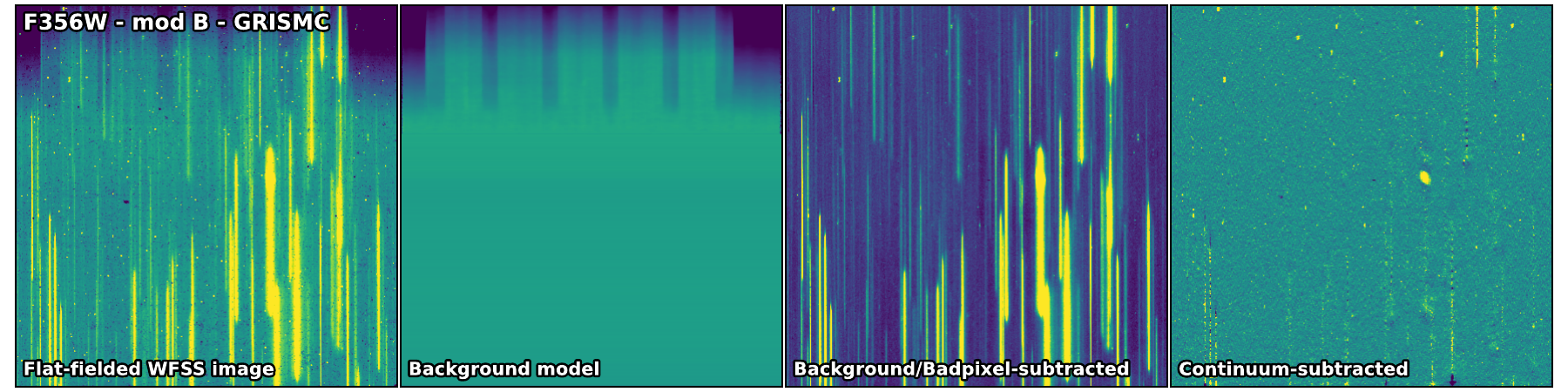}
\includegraphics[width=\linewidth]{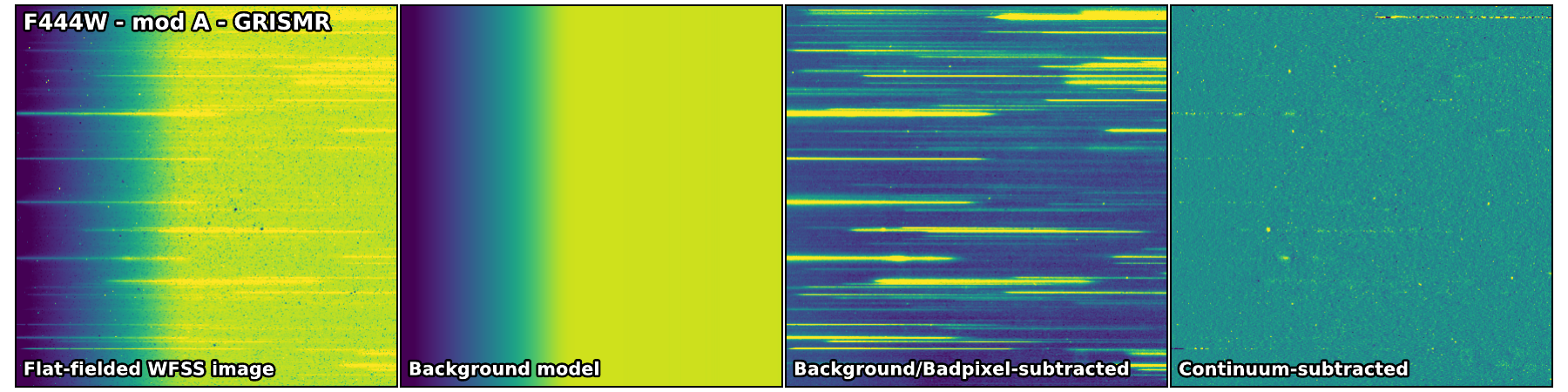}
\caption{Demonstration of NIRCam/WFSS Stage-2 processing for data taken with F356W Module B Grism C (top) and F444W Module A Grism R (bottom).
Our stage-2 processing includes flat-fielding (first column), modeled background subtraction and bad pixel (including hot pixels and cosmic rays) masking (second--third columns), and continuum subtraction (last column; See Section~\ref{ss:3c_grism}).
}
\label{fig:wfss_stg2}
\end{figure*}

\begin{figure*}[!t]
\centering
\includegraphics[width=\linewidth]{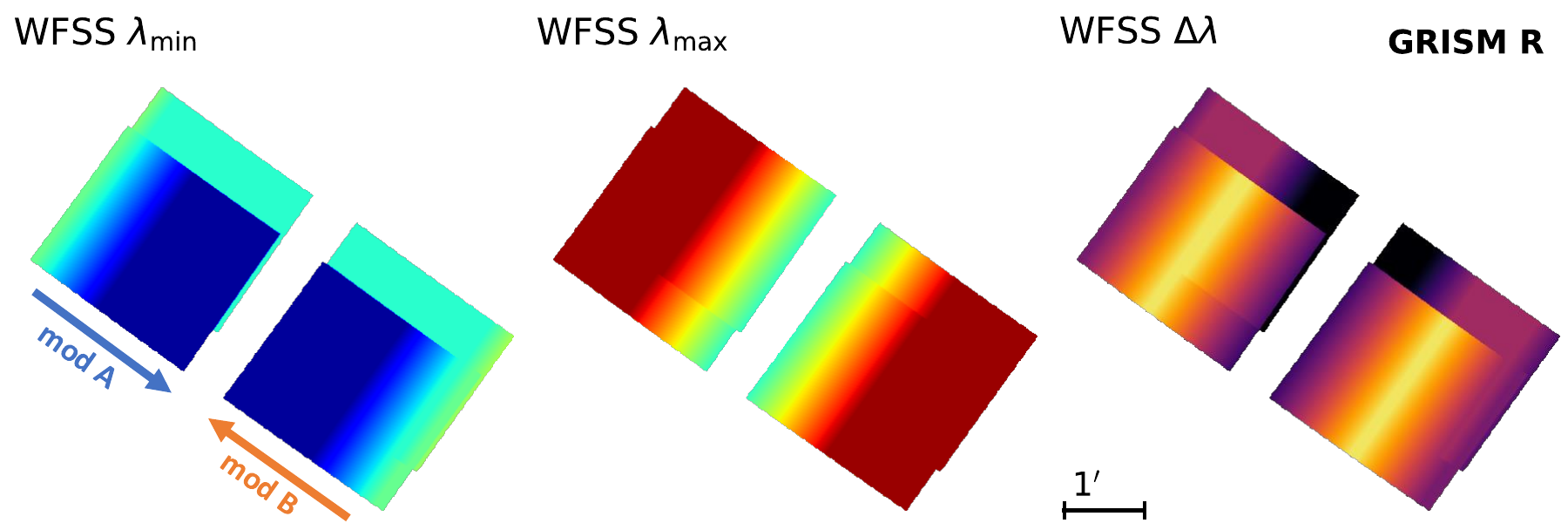}
\includegraphics[width=\linewidth]{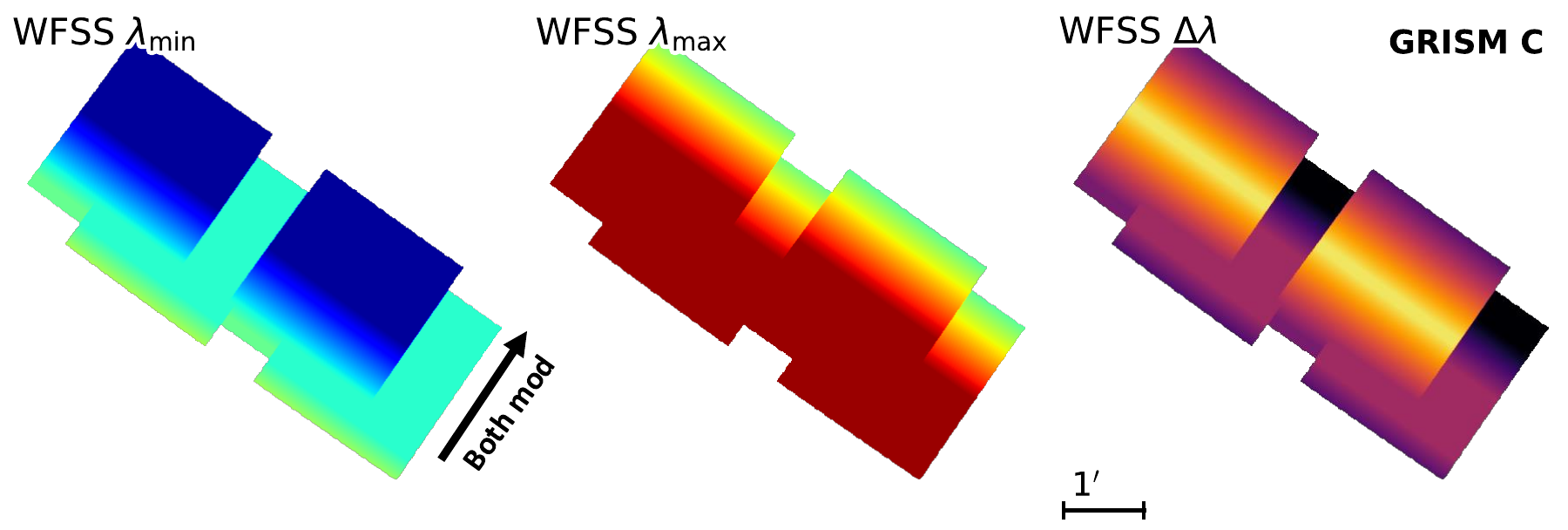}
\includegraphics[width=\linewidth]{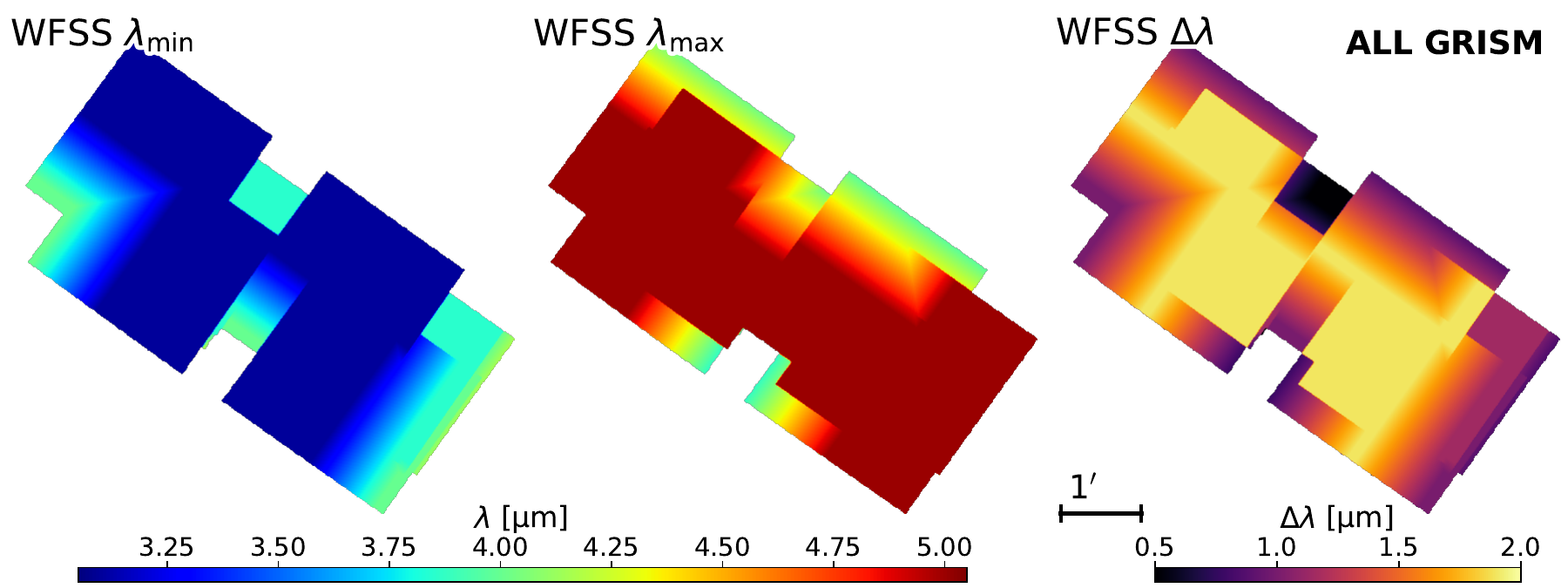}
\caption{NIRCam WFSS wavelength coverage map (minimum $\lambda_\mathrm{min}$, maximum $\lambda_\mathrm{max}$ and $\Delta \lambda = \lambda_\mathrm{max} - \lambda_\mathrm{min}$) by combining \sap\ EDR F356W and F444W WFSS observations.
Maps with Grism R (top), Grism C (middle) and all grisms (bottom) are shown.
Grism R / C dispersion directions on NIRCam module A and B are indicated by the arrows.
}
\label{fig:wfss_cov}
\end{figure*}

\begin{table}[!t]
\centering
\scriptsize
\begin{tabular}{cccccc}
\hline\hline
Filter & Grism & $t_\mathrm{exp}$ & $\lambda$ & Area & $5\sigma$ Depth \\
 & & [s] & [\micron] & [arcmin$^2$] & [$10^{-18}$\,\si{erg.s^{-1}.cm^{-2}}] \\
(1) & (2) & (3) & (4) & (5)  & (6)\\
\hline
F356W & R & 17007 & 3.2 &  6.0 & 1.15 \\
      &   &       & 3.4 &  6.9 & 0.83 \\
      &   &       & 3.6 &  7.8 & 0.69 \\
      &   &       & 3.8 &  8.7 & 0.55 \\
      &   &       & 4.0 &  9.0 & 1.26 \\
\hline
F356W & C & 17007 & 3.1 &  5.1 & 1.58 \\
      &   &       & 3.3 &  6.0 & 0.91 \\
      &   &       & 3.5 &  7.0 & 0.69 \\
      &   &       & 3.7 &  7.9 & 0.59 \\
      &   &       & 3.9 &  8.9 & 0.65 \\
\hline
F444W & R & 25510 & 3.9 & 12.0 & 0.98 \\
      &   &       & 4.1 & 11.9 & 0.71 \\
      &   &       & 4.3 & 10.8 & 0.76 \\
      &   &       & 4.5 &  9.6 & 0.85 \\
      &   &       & 4.7 &  8.5 & 0.87 \\
      &   &       & 4.9 &  7.3 & 0.93 \\
\hline
F444W & C & 25510 & 4.0 & 13.6 & 0.71 \\
      &   &       & 4.2 & 13.3 & 0.78 \\
      &   &       & 4.4 & 12.1 & 0.85 \\
      &   &       & 4.6 & 10.8 & 0.89 \\
      &   &       & 4.8 &  9.6 & 0.93 \\
      &   &       & 5.0 &  8.3 & 1.23 \\
\hline
\end{tabular}
\caption{Summary of \sap\ EDR grism observations. Column (1): NIRCam filter; (2): NIRCam grism; (3): Total exposure time in unit of second; (4): Wavelength in unit of \micron; (5): Effective area (in unit of arcmin$^2$) of FoV at given wavelengths; (6) Median $5\sigma$ depth (in unit of $10^{-18}$\,\si{erg.s^{-1}.cm^{-2}}) of unresolved lines at given wavelength, measured on module A (see Section~\ref{ss:4b_zsp} for details). The sensitivity with module B is shallower by $\sim 20$\%. }
\label{tab:02_grism}
\end{table}

\subsection{NIRCam WFSS Data Reduction}
\label{ss:3c_grism}
We reduced NIRCam WFSS data following the routine outlined by \citet{sun22b,sun23}, which includes a large number of customized steps and calibration different from those used by the standard \textsc{jwst} pipelines.
For transparency and reproducibility, the code and calibration files are publicly available\footnote{\url{https://github.com/fengwusun/nircam_grism}}, and the key steps are demonstrated in Figure~\ref{fig:wfss_stg2}.

Starting with the stage-1 data products from the standard \textsc{jwst} pipeline (Section~\ref{ss:3a_img}), we assigned the WCS to grism images and corrected for flat field using the STScI imaging flat-field data taken with the same filter.
% We generated the grism sky background model from a median stack of \sap\ and archival data taken with the same filter and grism.
The next step was grism sky background subtraction.
As shown in Figure~\ref{fig:wfss_stg2}, the 2D grism sky background $B(x,y)$ is almost independent of $y$ with Grism-R and thus easy to model. 
However, the 2D background is more complicated with Grism-C because of the coronagraph masks.
Therefore, we used deep F356W Grism-C data from GTO-4540 \citep[][private communication from F.\ Sun and D.\ Eisenstein]{eisenstein23b} to generate median-stacked 2D sky background models and subtract them off from the grism images.
We also further subtracted the sigma-clipped median along the row/column directions for Grism C/R images, respectively, to remove residuals of sky background and $1/f$ noise.
After this step, we further included a hot/bad-pixel masking step by rolling the grism image along the dispersion direction by 1\,pixel and identifying peaks in residual images at $>20\sigma$.
Finally, we also included the median-filtering technique introduced by \citet{kashino23} to produce a FITS extension that contains the emission-line only (\texttt{emline}).
To avoid over-subtraction around emission-line centroids, we first median-filtered the 2D grism image along the dispersion direction with a kernel size of 51 pixels, masking the central 9 pixels as ``holes''. 
We then identified pixels that deviated from zero at $>2\sigma$ from the median-filtering residual image, masked these pixels and again median-filtered the 2D grism image (similar to that in Section~\ref{ss:3a_img}) but with kernel size of 51 and 150 pixels through two iterations.
We also subtracted the median background along both row and column directions in the \texttt{emline} image.
% \clearpage

Accurate interpretation of NIRCam WFSS data relies on the spectral tracing ($y_s(x_0, y_0, x_s)$) and dispersion ($x_s(x_0, y_0, \lambda)$) functions obtained from calibration observations, where $(x_s, y_s)$ is the spectral pixel position on the 2D grism image at wavelength $\lambda$, and $(x_0, y_0)$ is the imaging position of the source on the detector \citep[i.e., the grism is not in the light path; see][]{sun22b, sun23}. 
With the latest calibration of NIRCam distortions \verb|jwst_nircam_distortion_0033.rmap| at the time of writing, the mapping between the sky coordinates of sources (R.A.\ and Decl.) to imaging position $(x_0, y_0)$ has been greatly improved.
Therefore, we have updated the calibrations of spectral tracing and distortion functions used by \citet{sun23} by including commissioning, Cycle-1 and 2 NIRCam WFSS data taken for SMP-LMC-58, a post-AGB star in the Large Magellanic Cloud (PID: 1076, 1479, 1480, 4449). 
The new calibrations were obtained in the same format as presented by \citet{sun23}.
The RMS accuracy of the tracing function is $\sim$0.07 native LW pixels ($\sim$\,0\farcs004), and the absolute wavelength calibration RMS is 0.1--0.2\,nm ($\Delta v \lesssim 15$\,\si{km.s^{-1}}).
Note that the accuracy of these calibrations relies on assumptions of stable internal alignment between LW and SW detectors, well-calibrated filter-introduced offsets and repeatability of filter and pupil wheels. 
All of these will need to be further assessed and calibrated through new observations.

With the new calibrations in hand and potential caveats on assumptions in mind, we first computed the astrometric offset of NIRCam LW grism images using the aforementioned source catalog (Section~\ref{ss:3b_cat}) and SW images taken simultaneously.
Because the primary NIRSpec observations have obtained proper target acquisition, the astrometric offsets are found to be small ($\lesssim0\farcs01$).
We further projected the source catalog (Section~\ref{ss:3b_cat}) to the astrometry-corrected grism WCS and applied the pick-off mirror masks, and therefore generated source catalogs of $(x_0, y_0)$ for each grism image, in which the sources could yield grism spectra.

We extracted grism spectra for all sources brighter than 29.8\,AB mag in the F356W band or 29.6\,AB mag in the F444W band.
Sources fainter than these limits are not expected to yield emission line detections at $\geq 5\sigma$ with the WFSS data.
2D spectra (both with and without continuum) were extracted, resampled and co-added (with sigma clipping) into a common wavelength grism of $\Delta \lambda = 1$\,nm, and the default height of the 2D spectra in the spatial direction is 31 native LW pixels (1\farcs95).
Spectra were flux calibrated based on grism sensitivity models characterized from Cycle-1 observations (PID: 1536, 1537, 1538).
We extracted 1D spectra from 2D spectra, both with and without continuum subtraction, using both boxcar aperture (height = $0\farcs31$) and the source profile (i.e., optimal extraction; \citealt{horne86}).
In practice, we cut out and collapsed the LW imaging data of the source along the dispersion direction, and use the resultant 1D profile in the spatial direction for extraction. 
However, the LW cutout images of certain sources might be subject to strong contamination from other nearby bright sources, and the resultant extraction profiles are incorrect. 
For these cases, we fit the central region of the 1D profile with a Gaussian model, and used the best-fit 1D Gaussian profile for optimal extraction.
Finally, for faint sources whose 1D collapsed profiles were noisy, we used the parametrized profile derived from source catalog (based on semi major/minor axes and positional angles).
Boxcar-extracted spectra were also corrected for aperture loss computed by these adopted spatial profile.

% Based on the grism tracing and dispersion function, we compute the wavelength coverage map 
The NIRCam WFSS FoV is wavelength-dependent because the offset between spectral pixel and source position $|x_s - x_0, y_s - y_0|$ is primarily controlled by wavelength and minimized as the so-called ``undeflected wavelength'' at $\sim3.94$\,\micron.
Based on the grism tracing and dispersion function, we compute the wavelength coverage map as Figure~\ref{fig:wfss_cov}.
The area with wavelength coverage $\Delta \lambda \geq 1$\,\micron\ is $\sim$\,12.0\,arcmin$^2$ with \sap\ EDR. 
The effective survey areas at 3.1--5.0\,\micron\ are also summarized in Table~\ref{tab:02_grism}.

\section{Value-added Data}
\label{sec:04_vad}

\subsection{Photometric Redshifts}
\label{ss:4a_zph}

We use \textsc{eazy} \citep{brammer08} to measure photometric redshifts (\zph) of sources in our EDR catalog, following a similar method as that adopted by the JADES team \citep{rieke23b,hainline24}.
We use the same spectral template sets adopted by \citet{hainline24}. % which has been proven effective for the selection of high-redshift galaxies. 
{We adopt the \textsc{eazy} error template ``\texttt{TEMPLATE\_ERROR.v2.0.zfourge}'' to account for wavelength-dependent uncertainties in the templates.}
We {explore the redshift range of $z=0.01-30$ with a redshift step of $\Delta z=0.01$}, derive the redshifts corresponding to the overall $\chi^2(z)$ minimum as our \zph, and use the output probability distribution $P(z) \propto \exp[-\chi^2(z) / 2]$ to compute the \zph\ uncertainties.
We use \texttt{circ1} photometry (without PSF matching) {and impose a minimum uncertainty of 5\% on the flux} for \textsc{eazy} SED fitting. 
This has been proven effective for the \zph\ selection of high-redshift galaxies \citep{hainline24}.
However, we also caution that the fits for large and red galaxies at slightly lower redshifts might be biased towards high \zph\ (see Section~\ref{ss:4b_zsp}).

\begin{figure*}[!t]
\centering
\includegraphics[width=\linewidth]{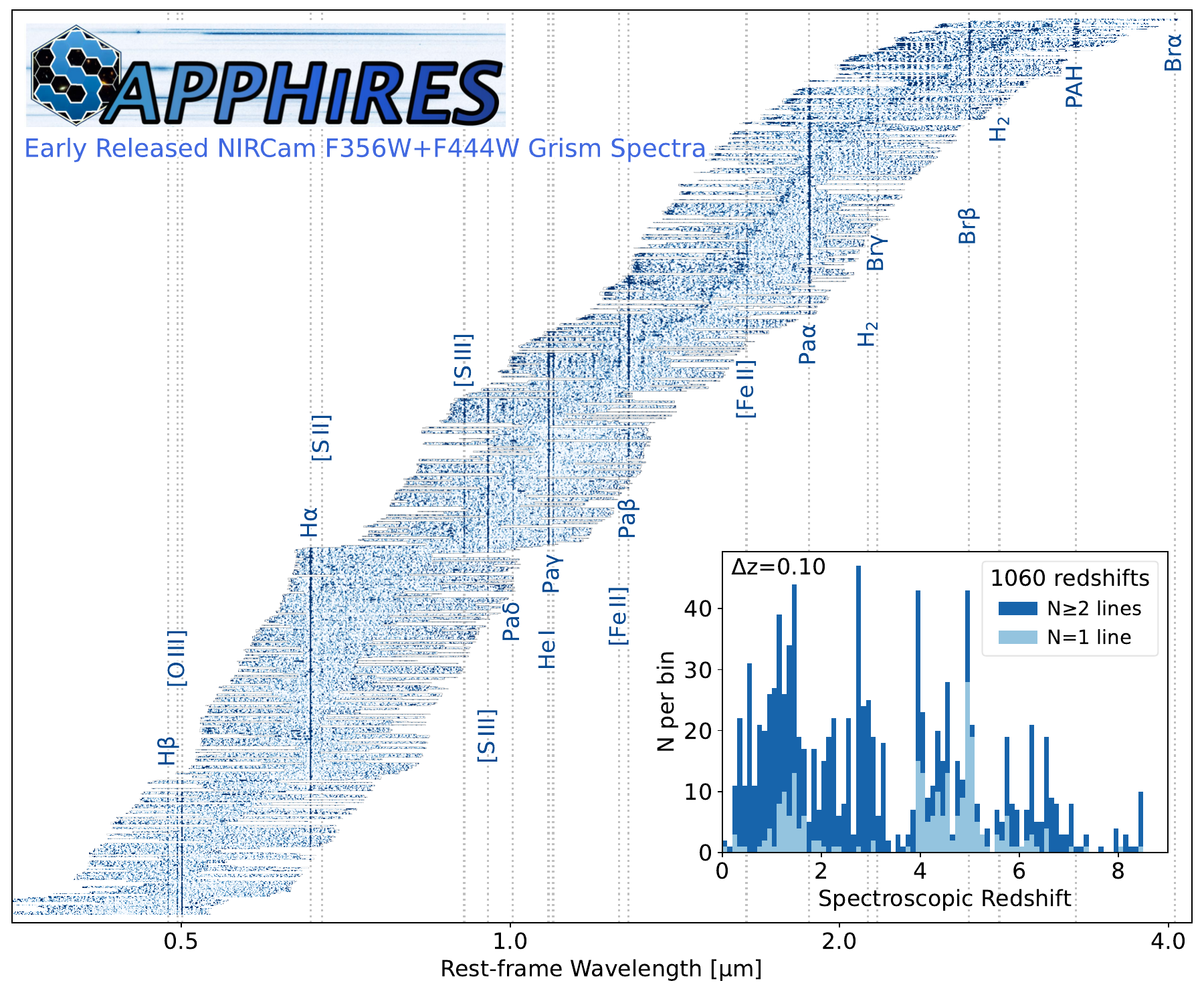}
\caption{\sap\ EDR emission-line spectra (F356W+F444W concatenated) stacked by the order of redshifts. 
Noticeable emission lines from \hb$+$\oiii\ at $z\sim8.5$ to \bra\ at $z\sim0$ are indicated with vertical dotted lines.
The inset panel shows the redshift histogram of confirmed emission line galaxies with a bin size $\Delta z = 0.10$.
}
\label{fig:spec-stack}    
\end{figure*}

\begin{figure}[!t]
\centering
\includegraphics[width=\linewidth]{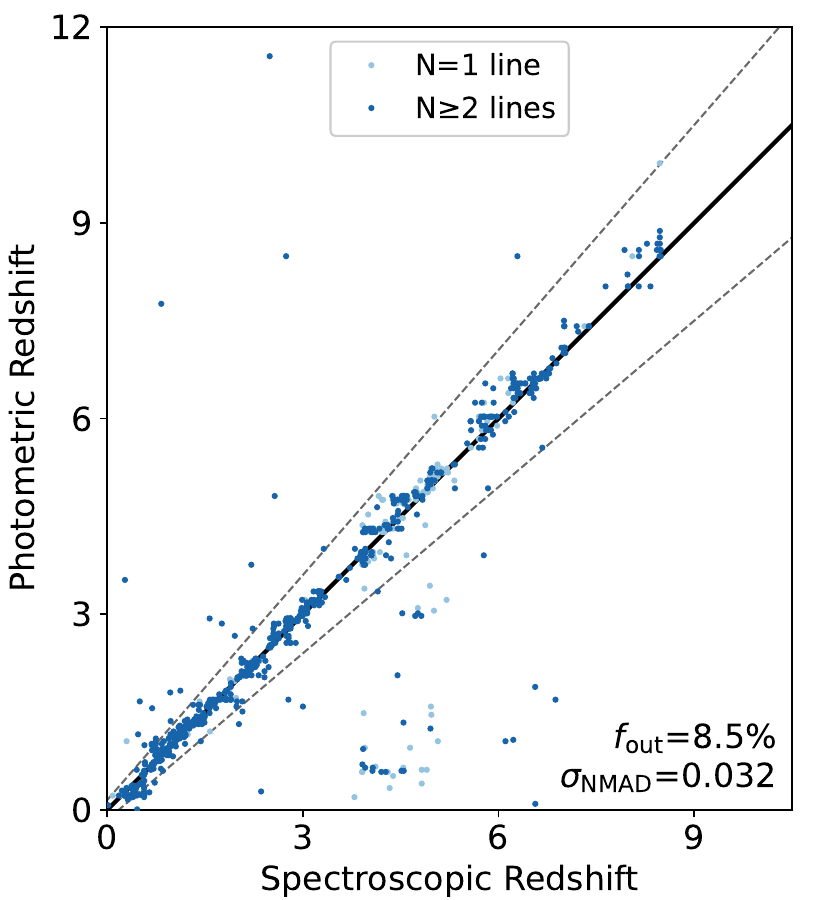}
\caption{Photometric redshifts compared with spectroscopic redshifts. The solid black line indicates \zph\,$=$\,\zsp, and dashed black lines indicate $|z_\mathrm{phot} - z_\mathrm{spec}| = 15\% (1 + z_\mathrm{spec})$ boundaries for catastrophic \zph\ outliers.
Sources with single and multiple line detections (\texttt{zconf=1-3} and \texttt{zconf=4-6}) are shown as light and dark blue dots, respectively.
}
\label{fig:zph_zsp}
\end{figure}

\subsection{Spectroscopic Redshifts}
\label{ss:4b_zsp}

We measure the spectroscopic redshifts (\zsp) based on the grism spectra and photometric redshifts using a semi-automatic algorithm presented in detail by \citet{Lin_GDN_HAE}.
As a brief summary, we first detect emission lines in both 1D and 2D spectra after continuum removal. 
If a detected emission line falls within the wavelength coverage of both Grism R and C spectra, we include only the lines that are simultaneously detected in both grisms to effectively remove contaminants. %\textblue{please check this sentence again}
We then apply a two-step cross-correlation algorithm between the detected emission lines and a series of emission-line templates for each source.
The emission-line template consists of frequently observed emission lines with NIRCam WFSS from \hb, \oiii\,$\lambda\lambda$4960,5008, \ha\ to the Brackett lines (\brg, \brb, \bra).
In the first step, we construct a model line spectrum with detected emission lines but at uniform fluxes and FHWMs to cross-correlate with templates at $z=0-10$ with a step of $\delta z = 0.1$.
The resulting series of correlation coefficients $C(z)$ peak at the redshifts where the lines in the template align with those in the observed spectra.
This coarse matching step saves computational resources and provides the possible redshift ranges for each galaxy.
In the second step, we set finer grids at $\delta z = 0.001$ across $z_l \pm 0.2$, where $z_l$ are likely redshifts with $C(z_l)$ computed in the first step.
The observed, continuum-removed spectrum is used (instead of the model line spectrum) and fit with the template spectra assuming realistic emission-line ratios at each $z$ grid. 
The resultant $\chi^2(z)$ is scaled by the probability distribution of the photometric redshift $P(z)$.

The best redshift solution of each source is selected based on $\chi^2(z)$ and visually inspected by multiple authors. 
Incorrect automatic redshift solutions are revised or removed from the visual inspections.
Because grism spectra of different sources can be often superposed, the whole redshift determination process is conducted in multiple iterations.
After each iteration the detected emission lines will be masked from the detector plane and thus the contaminated spectra in the next iteration.
% We also ensure that all sources with the same emission line detected at $\mathrm{S/N} \geq 5$ in both R and C grism spectra entering the redshift determination process.

After all iterations, we fit the grism spectra of each source around the best redshift solution to identify lines and measure the line centroids, fluxes and FWHMs.
Our line list includes (vacuum wavelength in \AA\ unless specified) \oii$\,\lambda$3727, \hb, \oiii\,$\lambda\lambda$4960, 5008, \ha, \sii\,$\lambda$6725, \siii\,$\lambda\lambda$9071, 9533, Paschen\,$\delta$ (\pad), \hei\,$\lambda$10833, \pag, \feii\,$\lambda$12570, \pab, \feii\,$\lambda$16440, \paa, \hei\,$\lambda$20592, H$_2$ 1--0 S(1)\,$\lambda$21223, Brackett $\gamma$ (\brg),  H$_2$ 1--0 Q(1)\,$\lambda$24072,  H$_2$ 1--0 Q(3)\,$\lambda$24243, \brb, H$_2$ 1--0 O(3)\,$\lambda$28032, PAH 3.29\,\micron,  Pfund\,8 (Pf8\,$\lambda$37405) and \bra, all of which have been detected with previous NIRCam WFSS observations.
The lines are fit with a Gaussian profile convolved with the grism line spread function (LSF; calibrated from observations; \citealt{Sun_LSF}).
Because the LSF wing could be somewhat over-subtracted in the continuum spectra with the median-filtering technique, we instead subtract the background and continuum in the calibrated 1D spectra by fitting a smoothing spline but masking the emission line region.
The emission line flux uncertainties are determined from covariance matrices of least-square fitting.
We only keep emission lines detected at $\mathrm{S/N}>3$ for our analyses.

The final redshift catalog of emission-line galaxies consist of sources that satisfy the following criteria:

\begin{enumerate}[topsep=0pt,itemsep=0ex,partopsep=1ex,parsep=1ex]
    \item The source has at least one emission line detected at $\mathrm{S/N} \geq 4$;
    \item The quadratically combined S/N of all identified emission lines $\sqrt{\sum_i(\mathrm{S/N})_i^2}$ is at $\geq 5$.
\end{enumerate}

This ensures the selection of single-line emitters at $\mathrm{S/N}\geq5$.
Note that the same spectral line observed with different filter--module--grism combinations is considered separately in the criteria above.
Redshifts are measured using the best-fit centroids of lines detected at $\mathrm{S/N}\geq4$, averaged with weights from their line S/N's.
Following \citet{Lin_GDN_HAE}, we provide scores (confidence levels, \texttt{zconf}) of \zsp\ at 1--6 according to the number of line detections and $\delta z = \zsp - \zph$:

\begin{itemize}[topsep=0pt,itemsep=0ex,partopsep=1ex,parsep=1ex]
\item \texttt{zconf=6}: At least two line detections and $|\delta z| \leq 0.2$;
\item \texttt{zconf=5}: $\geq 2$ lines and $0.2<|\delta z| \leq 1$;
\item \texttt{zconf=4}: $\geq 2$ lines and $|\delta z| > 1$;
\item \texttt{zconf=3}: One line detection and $|\delta z| \leq 0.2$;
\item \texttt{zconf=2}: One line and $0.2<|\delta z| \leq 1$;
\item \texttt{zconf=1}: One line and $|\delta z| > 1$;
\end{itemize}

In these criteria, the same emission line observed with different filter--module--grism combinations is considered as one line.
In total, we obtain \textblue{1060} redshifts of emission-line sources at $z \simeq 0 - 8.5$.
Figure~\ref{fig:spec-stack} displays the stack of spectra and redshift distribution of these sources.
Key emission lines are highlighted as validation of our redshift identification.
\textblue{819} sources have $N\geq 2$ line detections and \textblue{651} sources are at \texttt{zconf=6}.
The comparison of photometric redshifts and spectroscopic redshifts is shown in Figure~\ref{fig:zph_zsp}.
The fraction of sources with catastrophic \zph\ outliers, defined as $f_\mathrm{out} = |\delta z| / (1 + z_\mathrm{spec}) > 0.15$, is 8.5\%.
The scatter around the \zsp--\zph\ relation, i.e., normalized median absolute deviation, defined as $\sigma_\mathrm{NMAD} = 1.48 \times \mathrm{median}\big[\big|\delta z - \mathrm{median}(\delta z)\big| / (1 + z_\mathrm{spec})\big]$, is 0.032.
These values are intermediate between the statistics of JADES ($f_\mathrm{out} = 5\%$, $\sigma_\mathrm{NMAD}=0.024$; \citealt{rieke23b}) and UNCOVER ($f_\mathrm{out} = 11-17\%$, $\sigma_\mathrm{NMAD}=0.036-0.60$; \citealt{weaver24,price25}).
Because the parallel field was not covered with deep HST optical imaging such as the GOODS-S and Abell\,2744 field, most of the \zph\ outliers are galaxies at $z\sim5$ where the \lya-break is mistaken as the Balmer break in the bluest F070W band.
 
\begin{figure}[!t]
\centering
\includegraphics[width=\linewidth]{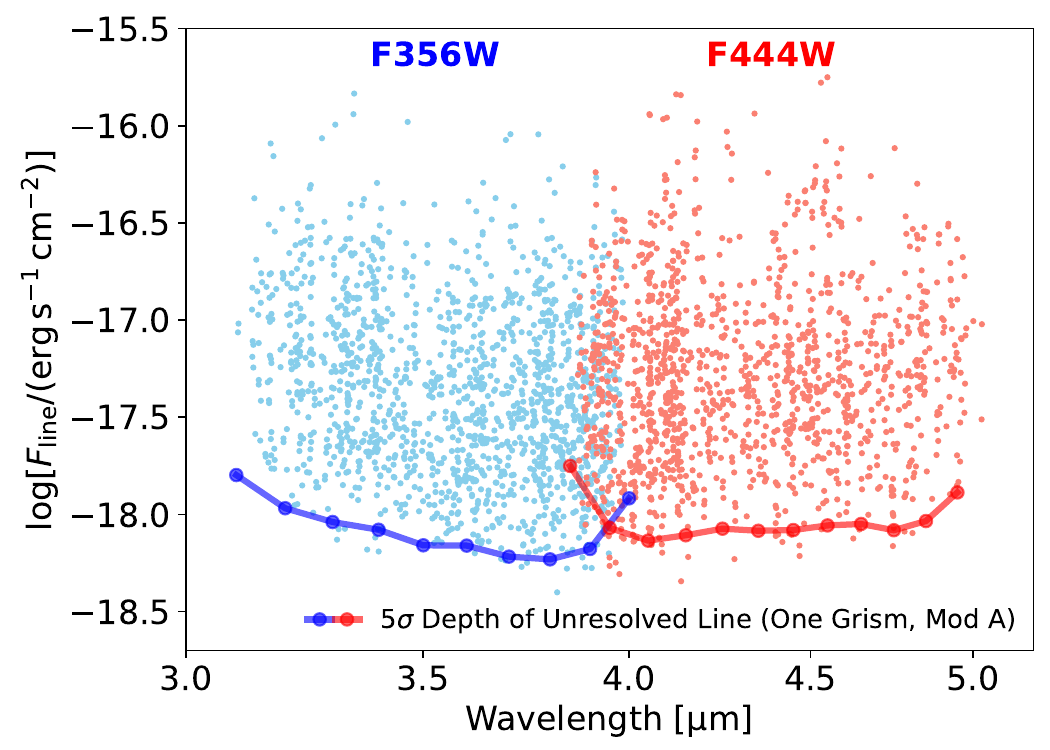}
\caption{Fluxes of emission lines detected at $\geq5\sigma$ versus wavelengths for lines in the F356W (blue) and F444W (red) band. 
The median $5\sigma$ depths for unresolved lines in module A are shown as the solid lines.
Note that spectral lines observed with different grisms are counted separately and not combined together.
}
\label{fig:linedepth}
\end{figure}

% highlighting the accuracy of our \zph\ measurements.
% Although this is higher than 
Such a large number of redshift confirmations over a small area of sky is a consequence of the great depth that \sap\ EDR achieved.
Figure~\ref{fig:linedepth} shows the fluxes versus wavelengths of emission lines detected at $\geq 5\sigma$.
We also compute the median $5\sigma$ depths of unresolved emission lines across the survey wavelengths in module A.
These are derived from the fluxes and uncertainties of compact emission-line sources with more than half of the fluxes within an $H=0\farcs31$ boxcar aperture and line FWHM smaller than 4\,nm, which are scaled to the observing condition (line FWHM and aperture loss) for unresolved lines at median exposure time.
The $5\sigma$ depths reach $6\times10^{-19}$\,\si{erg.s^{-1}.cm^{-2}} around 3.7\,\micron\ and $8\times10^{-19}$\,\si{erg.s^{-1}.cm^{-2}} around 4.2\,\micron\ (Table~\ref{tab:02_grism}).
These are slightly deeper than the NIRSpec observations of CEERS \citep{finkelstein25} and RUBIES \citep{degraaff24b} at the same wavelengths.
The emission line sensitivity with NIRCam module B is shallower by $\sim 20\%$.

% \textblue{add ETC prediction of Depth?\emoji{thinking-face}}

\begin{figure}[!t]
\centering
\includegraphics[width=\linewidth]{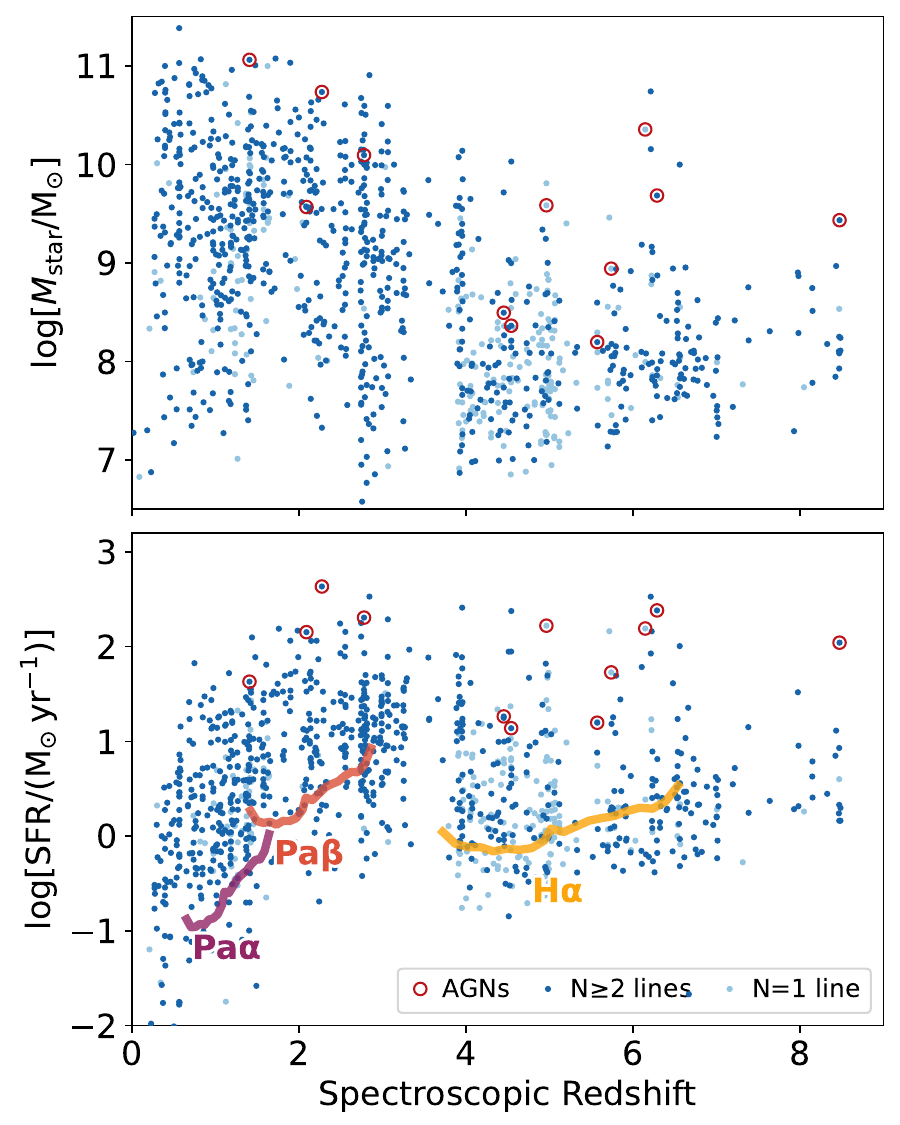}
\caption{Stellar masses (top) and SFR (bottom) versus spectroscopic redshifts for sources in EDR.
AGN and AGN candidates are highlighted by red circles, and neither their \mstar\ or SFR estimates are reliable.
Sources with single and multiple line detections are shown as light and dark blue dots, respectively. 
Nominal $5\sigma$ detection limits of \paa, \pab\ and \ha\ lines, converted into SFRs \citep[using the coefficients in][]{osterbrock06,ke12}, are shown in the bottom panel.
}
\label{fig:m_sfr_z}
\end{figure}

\begin{table}[!t]
\centering
\begin{tabular}{cccc}
\hline\hline
Line & $N$ & Redshift & EW (\AA) \\
(1) & (2) & (3) & (4) \\
\hline
\hb &  44 & 5.79--6.53--7.45 & 73--198--455 \\
\oiii\,$\lambda$4960 &  63 & 5.88--6.54--7.04 & 116--235--572 \\
\oiii\,$\lambda$5008 & 120 & 5.80--6.53--7.22 & 386--782--1389 \\
\ha & 291 & 3.98--4.72--5.58 & 345--702--1387 \\
\siii\,$\lambda$9533 & 158 & 2.55--2.89--3.27 & 45--99--188 \\
\hei\,$\lambda$10833 & 184 & 2.14--2.75--3.06 & 39--82--262 \\
\feii\,$\lambda$12570 &  87 & 1.75--2.13--2.72 & 8--24--75 \\
\pab & 153 & 1.61--2.14--2.74 & 22--50--127 \\
\feii\,$\lambda$16440 &  99 & 1.11--1.40--1.61 & 8--22--57 \\
\paa & 208 & 0.97--1.20--1.45 & 54--113--299 \\
\brg &  90 & 0.57--0.88--1.18 & 7--24--80 \\
\brb &  65 & 0.39--0.57--0.74 & 15--44--108 \\
\hline
% H2_2.12 &  74 & 0.58--0.94--1.14 & 7--20--64  \\
% H2_2.40 &  45 & 0.46--0.57--0.95 & 9--24--81  \\
% H2_2.42 &  46 & 0.43--0.58--0.97 & 5--17--74  \\
% H2_2.80 &  38 & 0.37--0.51--0.57 & 7--20--53  \\
% HeI_2.058 &  75 & 0.67--1.02--1.27 & 8--21--71  \\
% PAH &  18 & 0.27--0.36--0.40 & 42--144--373  \\
% S3_9071 & 121 & 2.63--2.99--3.32 & 17--40--139  \\
\end{tabular}
\caption{Statistics of key emission line detections in \sap\ EDR. Column (1): Name of emission line; (2) number of detections at $\geq3\sigma$; (3) 16--50--84th percentiles of redshift distribution; (4) 16--50--84th percentiles of rest-frame equivalent width (unit: \AA) distribution.
}
\label{tab:02_line_ew}
\end{table}

\subsection{Physical Properties}
\label{ss:4c_sed}

We use \textsc{Bagpipes} \citep{carnall18} to estimate the physical properties of spectroscopically confirmed galaxies by fitting their 13-band NIRCam SEDs.
Details of the modeling process, including priors, assumptions, and adopted models, can be found in \citet{Hsiao2023,Hsiao2024}, unless stated otherwise.
Briefly, to account for binary evolution, we adopt the Binary Population and Stellar Synthesis (BPASS) v2.2.1 templates \citep{Stanway2018}.
Nebular emission is included by reprocessing these stellar models through the photoionization code Cloudy \citep{Ferland2017}.
We assume a parametric star formation history following a delayed-$\tau$ model, the \citet{Calzetti2000} dust attenuation law allowing for extinction values in the range $0<A_{V}<8$ with a fixed $\eta=2$, and a \citet{Kroupa1993} initial mass function. 
We also include a photometric error floor of 5\%\ for SED modeling, to account for systematic errors arising from calibrations.

Figure~\ref{fig:m_sfr_z} shows the stellar mass and SFR versus redshift of spectroscopically confirmed sources in \sap\ EDR observations.
12 AGN and AGN candidates (including the so-called ``little red dots''; \citealt{matthee24}) at $z\simeq1.4-8.5$ are highlighted.
Because the accretion disk and broad-line region could contribute a significant fraction of light to the overall SEDs, and potentially even produce a Balmer break similar to that of an evolved stellar continuum \citep[e.g.,][]{inayoshi24,jix25}, the derived \mstar\ and SFR from \textsc{Bagpipes} can be unreliable.
Nevertheless, thanks to the great depth, the \sap\ EDR probes emission-line galaxies down to \mstar\,$\sim10^7$\,\msun\ out to $z\sim8$.

We observe increasing SFR as a function of redshift at $z\lesssim3$.
This can be explained by the redshift-dependent selection functions of \paa\ and \pab, key redshift indicators and star-formation tracers at the rest-frame near-IR.
With characteristic rest-frame equivalent widths of $\mathrm{EW}\sim10^2$\,\AA\ (see statistics in Table~\ref{tab:02_line_ew}) as observed with \sap, the detectability of these key near-IR lines quickly decreases toward higher redshift.
At $z>3.8$ where the strong \ha\ and subsequently \oiii\ lines enter the F356W bandwidth, the SFR selection limit is further improved to $\sim1$\,\smpy\ level.
We note that a substantial number of sources reside below the SFR limit converted from the \ha\ detection limit (\citealt{ke12}; assuming a \citealt{chabrier03} IMF).
This is because the SFR derived from the SED fitting is averaged over the last {10\,Myrs}, and many galaxies could be dominated by stellar populations younger than this timescale \citep[e.g.,][]{boyett24} given the presence of strong \ha\ and \oiii\ lines, resulting in $\mathrm{SFR}_\mathrm{10\,Myrs} < \mathrm{SFR}_\mathrm{H\alpha}$.
Alternatively, the elevated hydrogen ionizing photon production efficiency ($\xi_\mathrm{ion}$) at high redshifts can also be responsible for this result (see reviews by \citealt{robertson22}).

\section{Science Demonstration}
\label{sec:05_sci}

\begin{figure*}[!t]
\centering
\includegraphics[width=\linewidth]{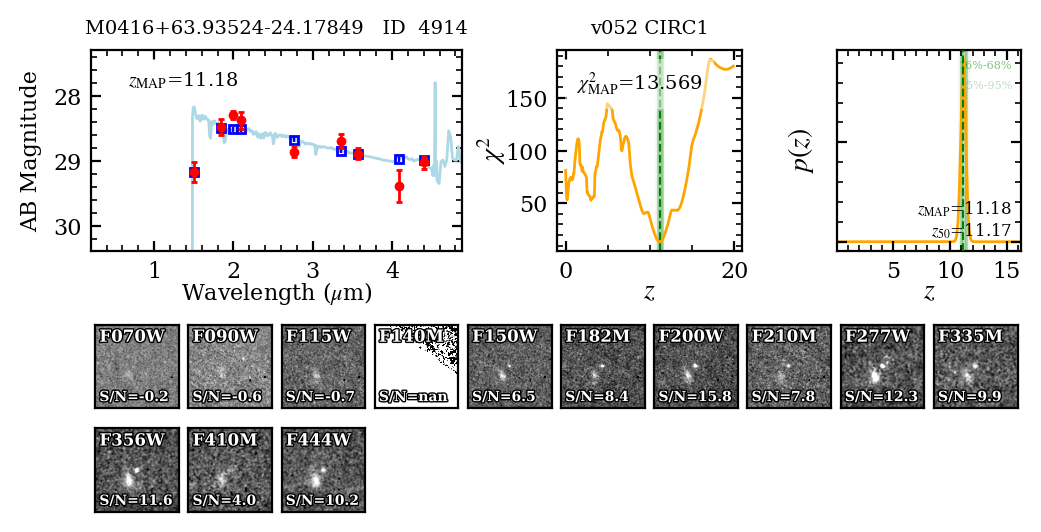}
\includegraphics[width=\linewidth]{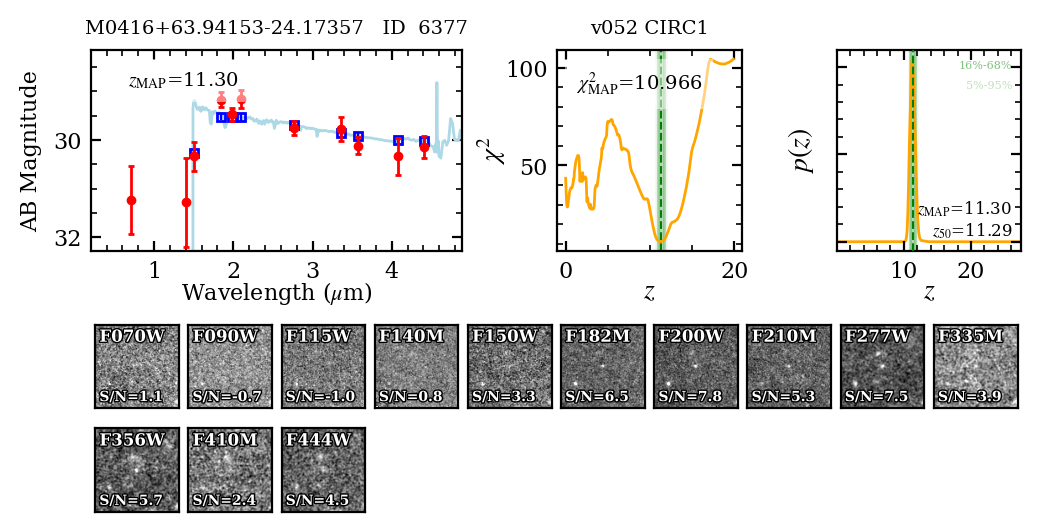}
\caption{Examples of two galaxy candidates at  $z>10$ identified with \sap\ EDR observations. 
For each source we show their 
(\romannumeral1, top left) observed SEDs (red circles with errorbars), best-fit SED models (light blue curve) and modeled flux densities in all bands (open blue squares); 
(\romannumeral2, top center) $\chi^2$ versus redshift of \textsc{eazy} fitting results; 
(\romannumeral3; top right) probability distribution of redshifts, highlighted with the redshift of maximal probability ($z_\mathrm{map}$, i.e., the \zph\ we adopt) and 5--16--50--84--95th percentiles of credible intervals;
(\romannumeral4, bottom) NIRCam image cutouts (\textblue{3\arcsec$\times$3\arcsec}) and S/N in each band.
\texttt{circ1} photometry is used.
}
\label{fig:z10}
\end{figure*}

In this section, we briefly present a few key science applications of \sap\ imaging and spectroscopic observations with early release data as examples.

\subsection{Galaxies at the Redshift Frontier ($z>10$)}
\label{ss:5a_z10}

\sap\ always obtains NIRCam imaging in the SW channel, and our design dedicates about 1/3 of the LW exposure time to direct imaging.
For the EDR observations, half of the exposure time was used for pure imaging observations over 13 bands at 0.6--5.0\,\micron, and thus offering great opportunities to photometrically select and characterize galaxies and AGNs over a broad redshift range.
We highlight the scientific applications of \sap\ as a deep pure-parallel imaging survey (i.e., similar to Cycle-1/2 NIRCam imaging pure-parallel programs PANORAMIC, \citealt{panoramic}; and BEACON, \citealt{beacon}) through the photometric selection of galaxies at the redshift frontier ($z>10$).

Figure~\ref{fig:z10} shows two examples of galaxy candidates at $z>10$ selected from the \sap\ EDR observations.
These two galaxies are selected with secure photometric redshifts at $z_\mathrm{phot} \sim 11$ that are tightly constrained by the partial F150W dropout.
The photometry in the F182M--F444W bands is consistent with the expectation from a blue rest-frame UV continuum.
The use of multiple medium-band filters such as F140M, F182M, F210M, F335M and F410M also greatly reduces the likelihood that the sources are $z<6$ interlopers due to strong Balmer break or emission lines, similar to the design of Cycle-2 program GO-3215 (JADES Origins Field; \citealt{eisenstein23b}).
At $\simeq29-30$ AB mag in the F356W and F444W bands, these two sources are too faint to yield spectral line detections with the grism spectra, but the brightnesses are consistent with those detected with previous JWST surveys in blank fields, e.g., JADES \citep{hainline24} and CEERS \citep{finkelstein23}.

Taking advantage of pure-parallel observations over multiple sightlines, the statistics of high-redshift galaxies from \sap\ will mitigate the impact from strong cosmic variance by pencil-beam surveys.
Based on the \citet{moster11} prescription, we anticipate that 4--5 observations with settings similar to \sap\ EDR will result in a cosmic variance level similar to that of JADES \citep{eisenstein23}, which surveys both GOODS fields over $\sim200$\,arcmin$^2$.
We note that other $z\gtrsim10$ galaxy candidates are also present in \sap\ EDR data and we anticipate presenting them in a future paper once the entire data have been acquired.

\begin{figure*}[!ht]
\centering
\includegraphics[width=\linewidth]{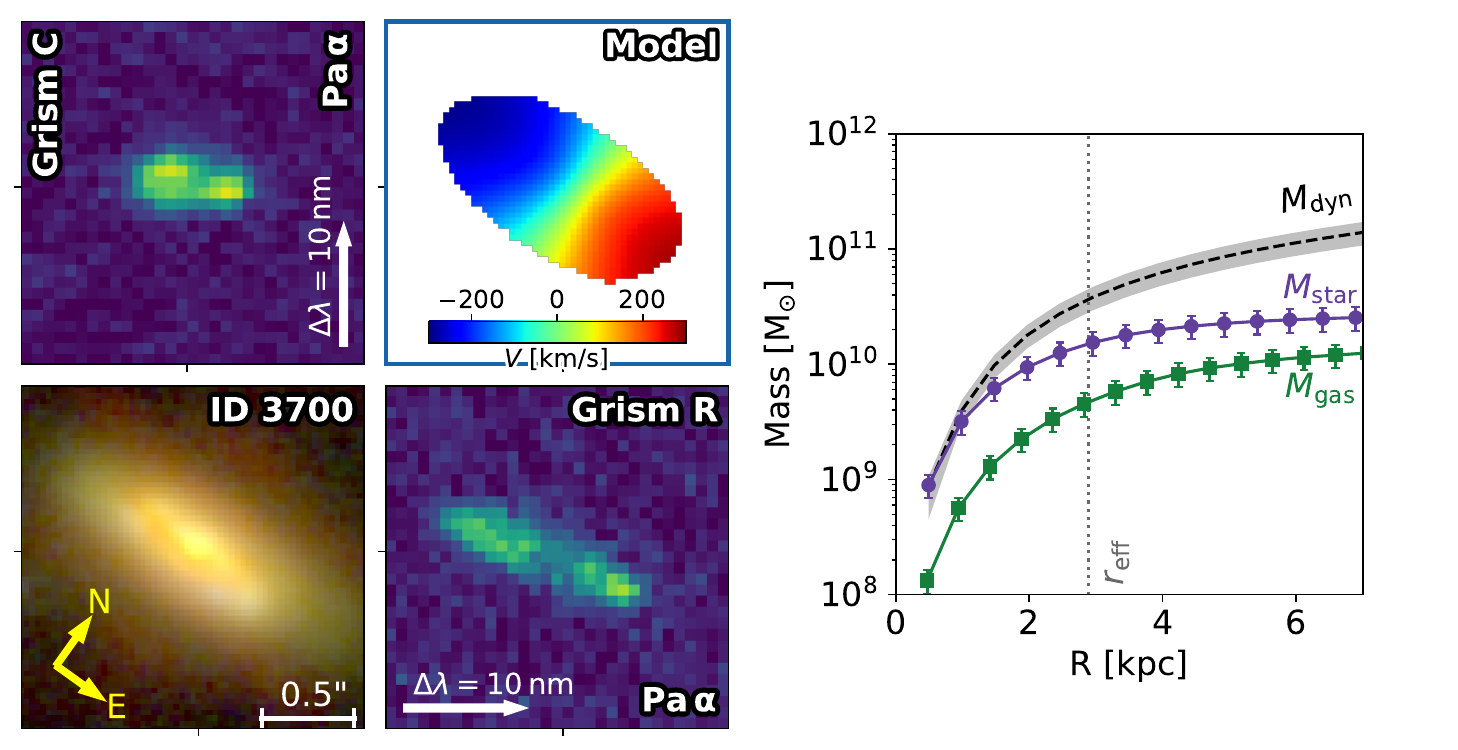}
\caption{\sap\ enables resolved gas kinematic measurements through deep grism spectroscopy at orthogonal dispersion directions. \textit{Left}: Example of a galaxy at $z=1.12$. The NIRCam false color image (red: F444W; green: F200W; blue: F090W) is shown in the lower-left corner, and \paa\ emission as observed with F444W Grism C (top-left) and Grism R (lower-right) are shown together with their dispersion directions (white arrows).
The difference in the emission-line morphology as observed with two grisms is a result of kinematic distortion.
The best-fit model of rotational disk is show in the top-right corner (see Section~\ref{ss:5b_kin}).
\textit{Right}: Radial dynamic (black), stellar (purple) and gas (green) mass profiles of the same galaxy, respectively, as modeled with \sap\ NIRCam imaging and WFSS observations (Section~\ref{ss:5b_kin}).
Half-light radius ($r_\mathrm{eff}$) is indicated by the vertical dotted gray line.
The galaxy mass is dominated by stars at $r\lesssim1$\,kpc and dark matter at $r > r_\mathrm{eff}$.
}
\label{fig:kinematic}
\end{figure*}

\subsection{Ionized Gas Kinematics}
\label{ss:5b_kin}

One unique aspect of \sap\ EDR observation is the use of both Grism R and C with both F356W and F444W filters.
Given the great depth that we have achieved in each filter-module-grism combination, this provides unique opportunities to study the kinematics of galaxies by forward modeling the high-S/N emission line maps (usually tracers of ionized gas and thus star formation) taken with grisms at $R\sim1500$ and orthogonal dispersion directions.

We demonstrate this scientific opportunity with an example in Figure~\ref{fig:kinematic}.
The source (ID 3700; R.A.\,=\,04$^h$15$^m$46.987$^s$, Decl.\,=\,--24\arcdeg10\arcmin50\farcs0) is a $z_\mathrm{spec}=1.123$ disk galaxy with \paa\ line detected at 3.98\,\micron\ through both grism R and C on module B.
The \paa\ emission-line maps observed with two grisms (projected into the same WCS) differ from each other, which is a result of kinematic distortion.
Previous grism kinematic studies using a single dispersion direction rely on strong priors of emission-line maps without kinematics \citep{nelson24,liz23,liuz25}, usually constructed from medium-band images or assumed from broad-band images.
The use of grism R and C together removes the dependence on such priors, enabling kinematic modeling from the grism images themselves.
In this example, we assume a simple arctangent rotation disk model as $V(r) = \frac{2}{\pi}V_\mathrm{rot}\,\mathrm{arctan}(r / R_v)$, where $V_\mathrm{rot}$ and $R_v$ are characteristic rotational velocity and radius, respectively.
For simplicity we do not include velocity dispersion in our model.
We perform Monte-Carlo Markov Chain (MCMC) modeling of the kinematics by matching the two distortion-corrected grism line maps.
The details of modeling for a larger sample of galaxies (and including bright \ha/\oiii\ emitters at $z\gtrsim6$) will be presented in a forthcoming paper from the collaboration.

The best-fit kinematic model is shown in Figure~\ref{fig:kinematic}.
Qualitatively, the \paa\ emission from the east side of the galaxy is stretched towards the redder direction with both 2D grism spectra, and the bluer direction for the west side.
This is consistent with our kinematic model.
We therefore derive the dynamic mass profile $M_\mathrm{dyn}(<r) = V^2 r / G$ applicable to thin disk.
In reality, we combine $V(r)$ and a baseline constant velocity dispersion ($\sigma = 60$\,\si{km.s^{-1}}; below the grism spectral resolution) quadratically to avoid $M_\mathrm{dyn}$ underestimate at the galaxy center.
The stellar mass profile $M_\mathrm{star}(r)$ is approximated from the F444W surface brightness profile, scaled to the stellar mass derived from SED modeling (Section~\ref{ss:4c_sed}) using the Kron-aperture photometry.
The gas mass profile $M_\mathrm{gas}(r)$ is estimated from the \paa\ surface brightness profile, as we convert the \paa\ surface brightness to SFR surface density and then gas surface density \citep{osterbrock06,ke12} according to the global Kennicutt-Schmidt relation \citep{kennicutt98b,kennicutt21}.
Taking potentially underestimated systematic errors into account, we adopt a 0.1-dex error floor for all radial profiles.

All profiles are shown in the right panel of Figure~\ref{fig:kinematic}.
The \paa\ emission is detected by the grisms out to $\sim7$\,kpc, i.e., $2.5\times$ of half-light radius ($r_\mathrm{eff}$) in the F444W band.
A sharp decline of the \paa\ surface brightness is seen at the centroid of the galaxy, indicating that the galaxy may undergo quenching from the inside out.
As is evident from the radial profiles, the galaxy mass in the inner $\lesssim1$\,kpc is dominated by the stellar component. 
The gas-to-stellar mass ratio increases as a function of radial distance from the galaxy center, and at $r > r_\mathrm{eff}$ the galaxy mass is dominated by the dark matter.

\begin{figure*}[!th]
\centering
\includegraphics[width=\linewidth]{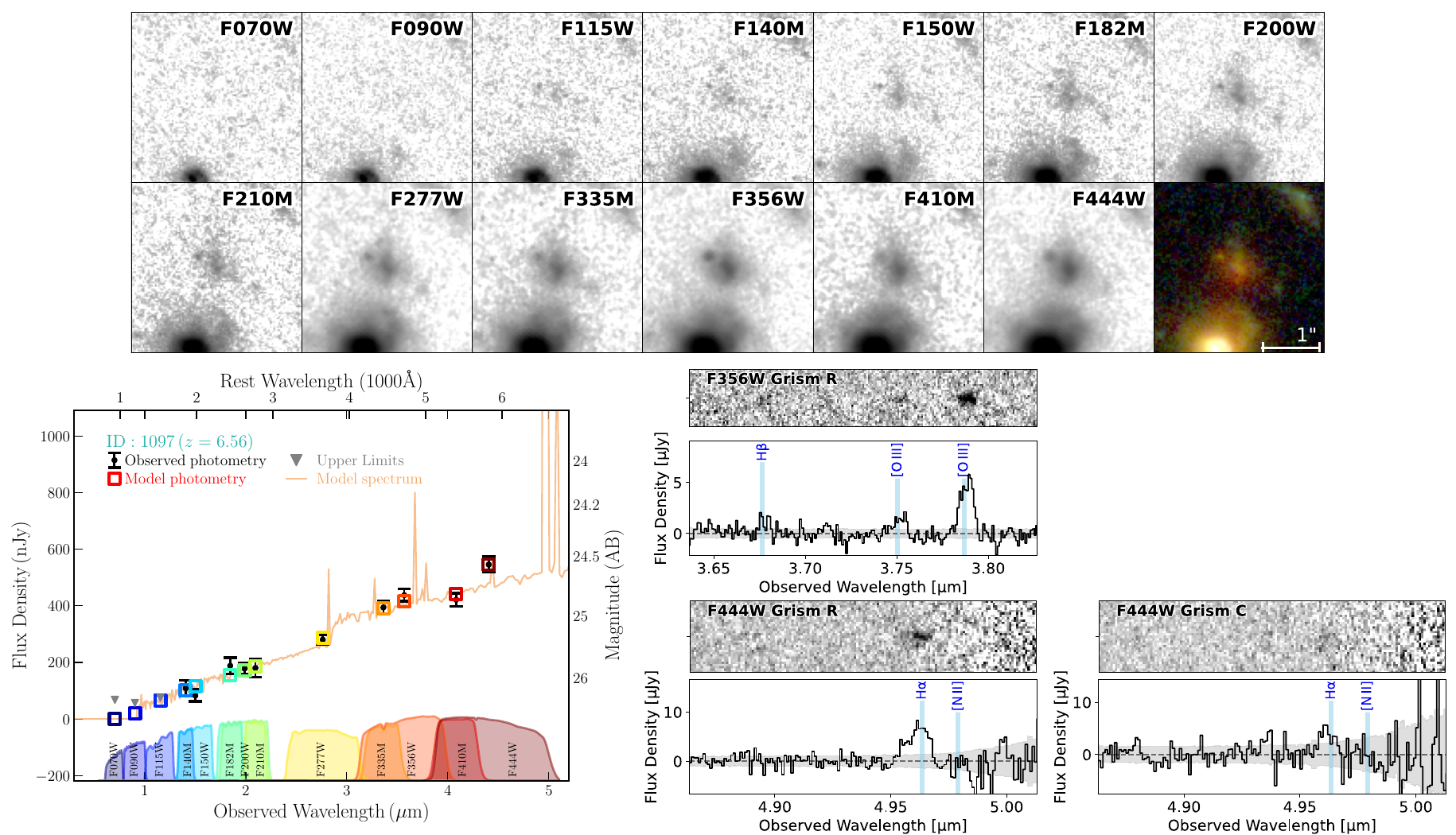}
\caption{A massive ($M_\mathrm{star}=10^{10.1\pm0.1}$\,\msun), dusty ($A_V=1.0\pm0.2$) star-forming galaxy confirmed at $z=6.56$ with \sap\ EDR (ID 1097).
\textit{Top}: 13-band NIRCam images of the galaxy. F444W-F200W-F090W RGB image is shown in the last panel. Image size is 3\arcsec$\times$3\arcsec.
\textit{Bottom-left}: Best-fit SED modeled with \textsc{Bagpipes}. 
\textit{Bottom-right}: NIRCam WFSS 2D and 1D spectra taken with F356W Grism R, F444W Grism R and C (background / continuum subtracted). 
The galaxy is out of the FoV of our F356W Grism C observations.
\hb, \oiii\,$\lambda\lambda$4960,5008 and \ha\ lines are detected and highlighted.
}
\label{fig:dsfg}
\end{figure*}

\begin{figure*}[!th]
\centering
\includegraphics[width=0.325\linewidth]{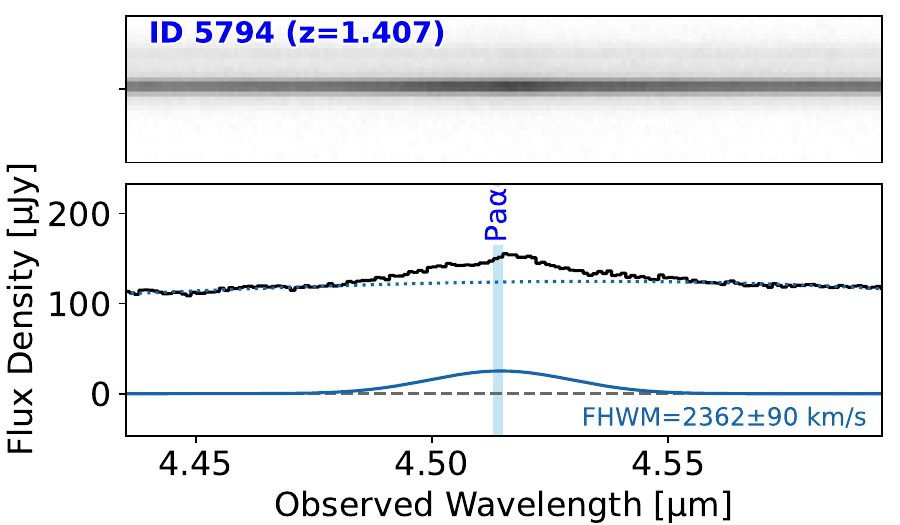}
\includegraphics[width=0.325\linewidth]{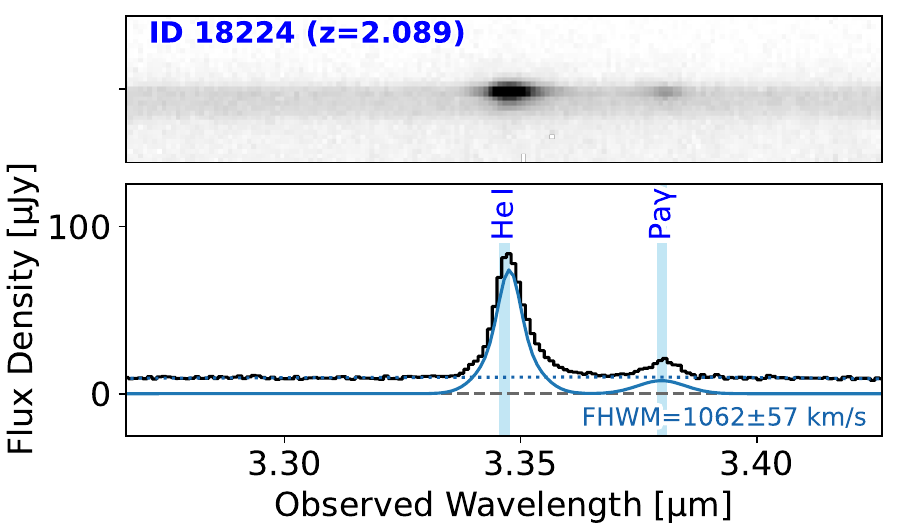}
\includegraphics[width=0.325\linewidth]{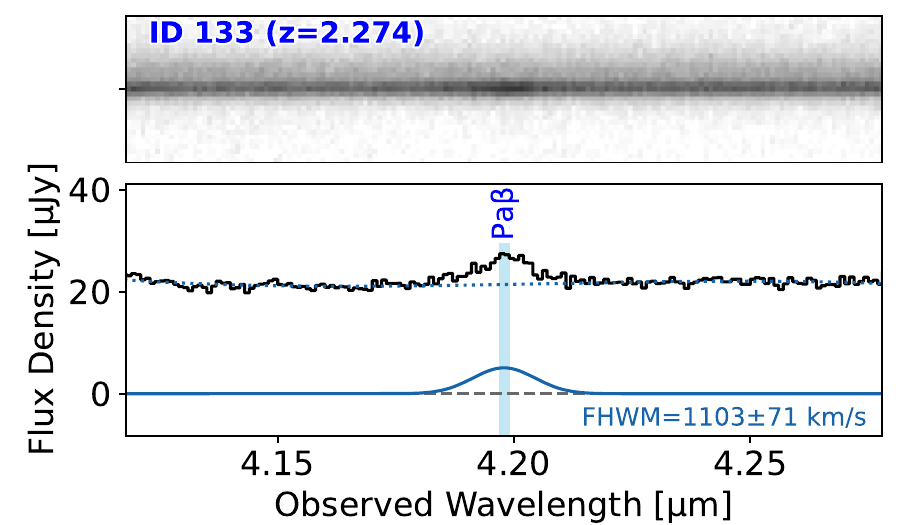}
\includegraphics[width=0.325\linewidth]{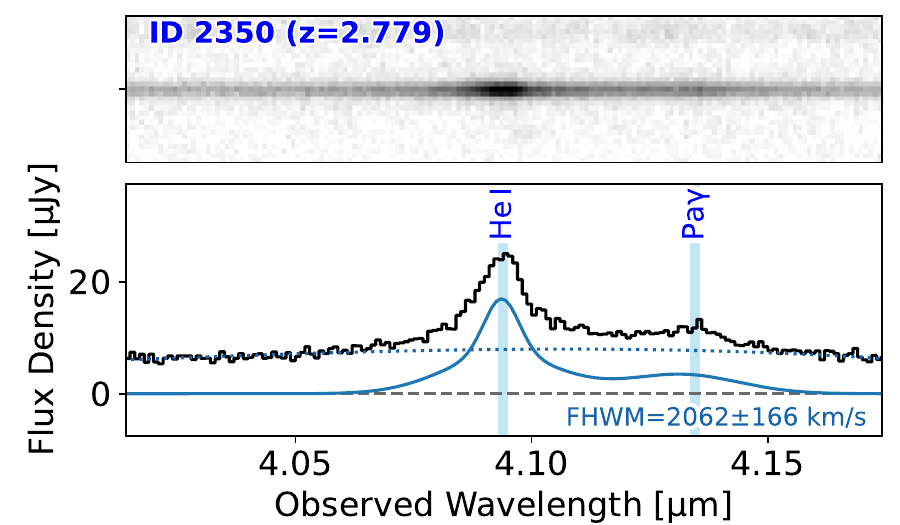}
\includegraphics[width=0.325\linewidth]{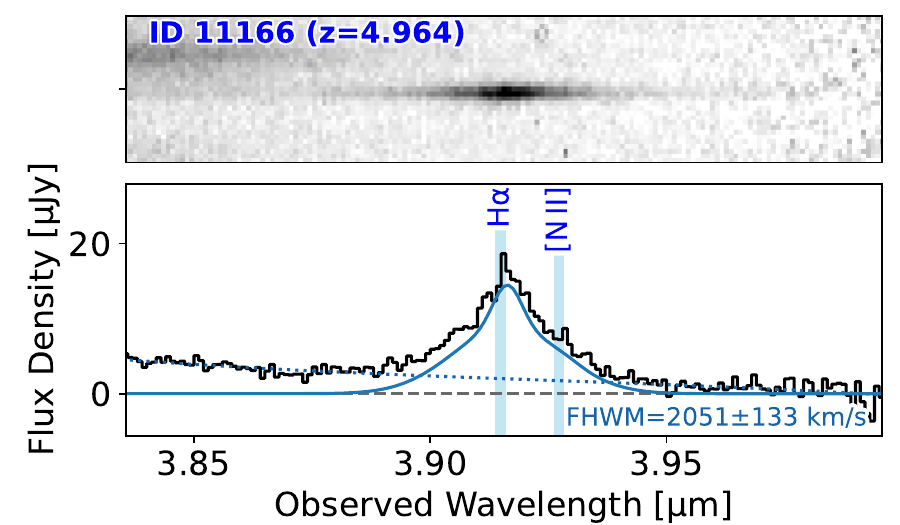}
\includegraphics[width=0.325\linewidth]{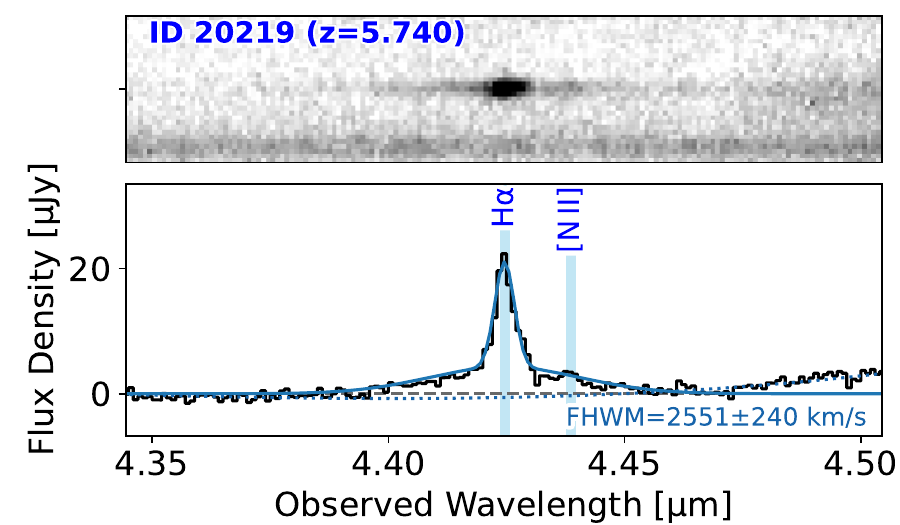}
\caption{Examples of six broad-line AGNs detected with \sap\ EDR data at $z\simeq1-6$.
In each plot we show their NIRCam WFSS 2D and 1D spectra without continuum removal.
Solid blue lines in the 1D spectral plot indicate the best-fit line profiles, and dashed blue lines indicated polynomial continuum / background model.
Source ID, redshift, broad-line FWHM and noticeable lines are indicated in each plot.
}
\label{fig:agn}
\end{figure*}

\subsection{Emission-line Galaxies and AGNs at the EoR}
\label{ss:5c_eml}

A key application of the NIRCam WFSS observations is to obtain a  highly complete luminosity-limited survey of emission-line galaxies and AGNs beyond the so-called ``cosmic noon'' and towards the EoR.
With a large survey area ($\sim500$\,arcmin$^2$ anticipated) and a variety of depths (0.5--20\,hr on-source integration), we expect to characterize the galaxy and AGN population at the EoR across a wide range of luminosities. 
Specifically, \sap\ is interested in (\romannumeral1) discovering and spectroscopically confirming the most luminous (massive) galaxies and AGNs at $z\simeq 4 - 9$ and potentially beyond through the wide survey area, and (\romannumeral2) characterizing the gas-phase metallicity and ionizing photon production of low-mass galaxies ($M_\mathrm{star} \lesssim 10^8$\,\msun) through flux-complete selection of \oiii, \ha\ and \hb\ lines with the deepest tier of the data, including this EDR.

We demonstrate the scientific opportunity for massive EoR galaxies with a brief case study of ID 1097, a luminous galaxy at $z_\mathrm{spec}=6.561$ showing strong dust attenuation (Figure~\ref{fig:dsfg}; R.A.\,=\,04$^h$15$^m$46.966$^s$, Decl.\,=\,--24\arcdeg11\arcmin33\farcs3).
This galaxy is spectroscopically confirmed with the detections of \hb, \oiii\,$\lambda\lambda$4960,5008 lines through F356W Grism R (4.7\,hours on-source), and \ha\ through F444W, both Grism R and C (7.1 and 2.3 hours on-source, respectively).
The spectral lines are broad, most likely due to morphological broadening as also seen and discussed with other NIRCam WFSS studies of high-redshift DSFGs \citep{sun24,sun25a}. 
The overall SED of the galaxy is red: the rest-frame UV continuum slope $\beta_\mathrm{UV}$ (such that $f_\nu \propto \lambda^{\beta_\mathrm{UV}}$) measured through F115W--F210M photometry is $\beta_\mathrm{UV} = -0.2 \pm 0.5$, and the Balmer decrement \ha/\hb\ is measured as $3.6\pm0.7$.
The red $\beta_\mathrm{UV}$ and less extreme Balmer decrement indicates a patchy dust geometry as frequently observed in high-redshift dusty star-forming galaxies \citep[e.g.,][]{penner12, oteo13, casey14b, sun24}, where the emission lines may mostly originate from less obscured regions compared with that for stellar continuum.
The detailed spatially resolved study will be presented with an upcoming paper.

The SFR measured from the UV light ($\mathrm{SFR}_\mathrm{UV}=4\pm1$\,\smpy) is much smaller than that measured from \ha\ ($\mathrm{SFR}_\mathrm{H\alpha} = 18\pm 2$\,\smpy; uncorrected for dust).
Therefore, we conclude that the unobscured SFR at the rest-frame UV is $\lesssim 10$\%\ of the total SFR (inferred from IR Excess, IRX--$\beta_\mathrm{UV}$ relation; e.g., \citealt{meurer99}), which is mostly obscured by dust.
This is also reflected by the best-fit total SFR from \textsc{Bagpipes} SED modeling $\mathrm{SFR}_\mathrm{SED} = 101_{-44}^{+133}$\,\smpy.
Such a large SFR at $z\sim7$ is often required for the production of most massive quiescent galaxies seen at $z\gtrsim4$ according to their star-formation history analyses \citep[e.g.,][]{carnall23b, carnall23a, degraff24, glazebrook24, turner25}. 
We also note that SED modeling suggests a stellar mass of {${M}_\mathrm{star} = 10^{10.0 \pm 0.1}$\,\msun}\ for the system.
Therefore, we also conclude that ID 1097 is representative of the most massive galaxies at $z>6$ expected in our EDR survey volume given the present measurements of galaxy stellar mass function \citep{shuntov24,weibel24a}.

NIRCam WFSS is also highly efficient for the selection of broad-line AGN across a wide redshift range \citep[e.g.,][]{linx24,matthee24,zhuangm24,suny25}.
Figure~\ref{fig:agn} displays six of the broad-line AGNs detected within \sap\ EDR at $z_\mathrm{spec} \simeq 1 - 6$.
Among them, ID 11166 is particularly interesting because (\romannumeral1) the \ha\ emission from the AGN and the host galaxy at 3.91\,\micron\ are detected with both Grism R and C observations in both F356W and F444W bands, and (\romannumeral2) it resides in a galaxy overdensity as shown in the redshift histogram of Figure~\ref{fig:spec-stack}.
The highest-redshift broad-line AGN that we have identified with \sap\ EDR is at $z_\mathrm{spec} = 8.479$, which is also embedded in a galaxy overdensity  \citep[see details in][]{Fudamoto_OD}.
By fitting the 1D spectra of sources in Figure~\ref{fig:agn} with polynomial continuum models and Gaussian profiles, broad \paa, \pab, \pag, \hei\,$\lambda$10833 and \ha\ lines are detected with FWHM above 1000\,\si{km.s^{-1}}.
The detailed characterization of broad-line AGNs, including their redshift distribution, number densities, properties of black holes and host galaxies as well as their clustering properties will be foci of future papers from the collaboration.

\section{EDR Data Products}
\label{sec:06_data}

\subsection{Data Products and Access}
\label{ss:6a_data}

The \sap\ EDR data products being released with this paper include the following items:

\begin{enumerate}[topsep=0pt,itemsep=0ex,partopsep=1ex,parsep=1ex]

\item 13-band NIRCam image mosaics (Section~\ref{ss:3a_img});

\item Photometric catalogs of 22107 sources with photometric redshifts measured with \textsc{eazy} (Section~\ref{ss:3b_cat} and \ref{ss:4a_zph});

\item Spectroscopic redshift catalog of \textblue{1060} sources with physical properties modeled by \textsc{Bagpipes} (Section~\ref{ss:4b_zsp} and \ref{ss:4c_sed});

\item 1D, 2D NIRCam grism spectra of \textblue{1060} sources with grism redshift measurements (Section~\ref{ss:3c_grism} and \ref{ss:4b_zsp}).

\end{enumerate}

These data are available at SAPPHIRES team website\footnote{\url{https://jwst-sapphires.github.io/}}, and will be made available at MAST as High-Level Science Products (HLSP) once the preparation is ready. %\footnote{ \url{https://archive.stsci.edu/hlsp/sapphires}}.}
This site also includes detailed documentation of the data structure and catalog contents. 
Further, following JADES \citep{eisenstein23,rieke23b} and NEXUS \citep{sheny24,zhuangm24} practice, we also provide a link to our \textsc{FITSmap} online visualization tool (\citealt{hausen22}; see our team website).
Users can pan and zoom in NIRCam data taken with multiple filters, and switch on different layers to inspect the associated catalogs, plots of \textsc{eazy} SED and NIRCam grism spectra.

Further high-order data products (e.g., PSF-matched images and photometry, catalogs of emission-line fluxes, spectra of other sources) and data products of other observations taken with \sap\ will be the subject of our further data releases.
Pre-release data products are also available from the \sap\ team upon reasonable request.
We also note that \sap\ is a Treasury program without proprietary period and all raw and calibrated data (processed through the default STScI pipeline) are available on MAST.

\subsection{Caveats}
\label{ss:6b_caveat}

We present a few caveats on the EDR products and users should be aware of these known defects / artifacts. 

The NIRCam imaging and WFSS data were taken in parallel to JWST-GO-4750 NIRSpec MOS observations with the default three-shutter nodding pattern. 
Therefore, the bad pixel and cosmic ray rejection (especially in imaging data) may remain unreliable when the number of exposures is $\leq3$.
Users are encouraged to visually inspect the data for sources that are low in S/N or only detected in a limited number of bands.

As presented in Section~\ref{ss:3b_cat}, the photometric catalog overshreds large galaxies to enhance the survey completeness of emission-line galaxies at higher redshifts.
Galaxies with substructures (e.g., spiral arms and clumps) are therefore usually deblended into multiple entries in the photometric catalog (and sometimes the spectroscopic catalog), resulting in duplication and potentially imprecise measurements of fluxes and physical properties as a consequence.
Given the successful discovery of JADES-GS-z14-0 (a luminous galaxy at $z=14.2$ that is heavily blended with a foreground galaxy; \citealt{hainline24,robertson24,carniani24a}) and our experience with NIRCam WFSS data analyses, we opt to adopt such a deblending routine throughout the \sap\ survey.
Users are therefore encouraged to examine the measurements of large galaxies to see if they are over-shredded in our catalogs.

The FoV of the NIRCam WFSS mode is wavelength-dependent (Figure~\ref{fig:wfss_cov}).
Non-detection of emission lines does not necessarily mean the absence of strong emission line within the F356W or F444W bandwidths, as the lines could be outside of the effective spatial and wavelength coverage or simply omitted through our semi-automatic redshift determination routine (Section~\ref{ss:4b_zsp}).
Users are encouraged to inspect the spectra for sources without expected emission-line detections.

Because the NIRCam WFSS obtains spectra from all sources entering its effective FoV, spectral overlapping and contamination are inevitable for all NIRCam WFSS surveys.
The mitigation of continuum contamination requires proper forward modeling \citep[e.g.,][]{pirzkal17,Smith_IceAge}, which is currently limited to bright continuum sources.
Strong continuum contamination could affect the detectability of faint sources especially when the expected line is around the edge of the filter transmission curve.
Emission-line contamination could lead to incorrect spectroscopic redshift determinations, and sometimes false detection of faint emission lines that artificially boost the line fluxes and redshift confidence level.
Users are encouraged to visually inspect the spectra (1D and 2D) used for analyses, and mask or subtract the contaminants if present.

A median filtering technique is used for emission-line identification (Section~\ref{ss:3c_grism}). 
Although masks have been employed for strong lines, the broad wing of emission lines may be still slightly oversubtracted in these data products.
Emission lines in NIRCam grism spectra could be broadened by a variety of reasons, e.g., intrinsic broadening from the PSF and LSF \citep{Sun_LSF},  extended morphology and kinematics (e.g., presence of outflows or broad-line regions).
For best accuracy, users are therefore recommended to perform line flux measurements on data without the default continuum subtraction, and remove the continuum through methods such as smoothing spline interpolations or polynomial fitting of the spectra.

The \textsc{eazy} photometric redshift measurements are subject to a 8.5\%\ catastrophic outlier fraction as discussed in Section~\ref{ss:4a_zph}.
In addition to this, stars and brown dwarfs are effectively treated as galaxies through \textsc{eazy} modeling and the \zph\ could be obviously wrong ($z_\mathrm{phot} \gtrsim 6$; e.g., \citealt{hainline24b}).
In the physical SED modeling with \textsc{Bagpipes}, we adopt the default Kron-aperture photometry, which might lead to low S/N and thus poorly constrained SED models for intrinsically faint sources especially at high redshifts. 
Users are recommended to compare with SED modeling results using smaller apertures (e.g., \texttt{kron\_s} or \texttt{circ2}).
Finally, the SEDs of AGN (and AGN candidates) modeled with \textsc{Bagpipes} through galaxy stellar population synthesis can also be highly unreliable.
These artifacts shall be taken into account in further data analyses.

\section{Summary}
\label{sec:07_sum}

We present the Early Data Release of the Slitless Areal Pure-Parallel HIgh-Redshift Emission Survey, \sap\ (JWST-GO-6434).
This is the initial data release for this JWST Cycle-3 Treasury imaging and spectroscopic program utilizing the NIRCam grism spectroscopic mode in pure parallel, totaling a telescope charged time of 709 hours (557 hours of exposures).
The main content of this paper and EDR is summarized as follows:

\begin{enumerate}[topsep=0pt,itemsep=0ex,partopsep=1ex,parsep=1ex]
\item NIRCam imaging and WFSS data were taken as pure parallel to JWST-GO-4750 (PI: Nakajima), which obtained NIRSpec MOS observations in the Frontier Field galaxy cluster MACS\,J0416.1--2403.
We obtained NIRCam data as parallel observations with a dual-channel exposure time of 47.2 hours, and our footprint is outside of the Frontier Field with no archival JWST or HST data included for our data analyses.

\item We obtained deep 13-band NIRCam imaging data (from F070W to F444W, including five medium bands) over an area of $\sim16$\,arcmin$^2$.
The median $5\sigma$ depth of point source (measured with $r=0\farcs15$ aperture) reaches 28.4--29.5\,AB mag across all filters.
22107 entries of sources are photometrically catalogued from the images.

\item We obtained deep NIRCam WFSS data with the F356W and F444W filters at orthogonal dispersion directions (through Grism R and C).
The $5\sigma$ depths for unresolved emission lines reach $6\times10^{-19}$\,\si{erg.s^{-1}.cm^{-2}} around 3.7\,\micron\ and $8\times10^{-19}$\,\si{erg.s^{-1}.cm^{-2}} around 4.2\,\micron\ (both through NIRCam module A).

\item We measure photometric redshifts of cataloged sources through \textsc{eazy} SED modeling, and obtain 
\textblue{1060} spectroscopic redshifts at $z\simeq0-8.5$ through a semi-automatic algorithm with human inspection.
We measure the stellar mass, SFR, dust attenuation and ages of redshift-confirmed sources through \textsc{Bagpipes} SED modeling.

\item We demonstrate a few scientific applications of \sap\ EDR by presenting (\romannumeral1) $z>10$ galaxy candidates, 
(\romannumeral2) ionized gas kinematics and mass profiles of a disk galaxy at $z=1.123$, 
(\romannumeral3) a massive dusty star-forming galaxy at $z=6.561$,
and (\romannumeral4) broad-line AGNs selected across $z\simeq1-6$.

\item We describe how to access the EDR including NIRCam image mosaics, photometric catalog (including derive \zph), spectroscopic redshift catalog (including derived physical properties) and spectra of redshift-confirmed galaxies and AGNs.
We also discuss caveats for the released data products.

\end{enumerate}

The data released through this paper represent only only $\sim10$\% of the observations that will be ultimately acquired with the \sap\ survey.
The EDR is aimed to enhance the transparency and visibility of \sap\ as one of the JWST Treasury programs approved in Cycle 3, and attract interests from the world-wide community of extragalactic astronomy and potential JWST NIRCam WFSS users.
Future releases of \sap\ imaging-spectroscopic data and value-added data products will be presented once the data acquisition is complete.

\section*{Acknowledgment}

We thank the program coordinator of \sap, Shelly Meyett, for the extremely helpful, timely and continuous support offered to our program.
We thank STScI staff including (but not limit to) Martha Boyer, Mario Gennaro, Nor Pirzkal and John Stansberry for their work in calibrating the telescope and instrument and making NIRCam WFSS pure parallel observations possible.
F.S.\ and E.E.\ thank Jeyhan Kartaltepe, Marc Rafelski (PIs of POPPIES, JWST Cycle-3 GO-5398) and Christina Williams (PI of PANORAMIC; JWST Cycle-1 GO-2514) for very helpful discussions, which are inspiring for our program design and implementation.
F.S.\ and E.E.\ thank Karl Misselt for his work in maintaining our computational resources.

F.S., J.M.H., E.E., D.J.E., C.N.A.W.\ and Y.Z.\ acknowledge JWST/NIRCam contract to the University of Arizona NAS5-02015. 
A.J.B.\ acknowledges funding from the ``FirstGalaxies” Advanced Grant from the European Research Council (ERC) under the European Union’s Horizon 2020 research and innovation program (Grant agreement No. 789056).
F.W. acknowledges support from NSF award AST-2513040.
This work is based on observations made with the NASA/ESA/CSA James Webb Space Telescope. The data were obtained from the Mikulski Archive for Space Telescopes at the Space Telescope Science Institute, which is operated by the Association of Universities for Research in Astronomy, Inc., under NASA contract NAS 5-03127 for JWST. These observations are associated with program \#6434.
Support for program \#6434 was provided by NASA through a grant from the Space Telescope Science Institute, which is operated by the Association of Universities for Research in Astronomy, Inc., under NASA contract NAS 5-03127.

\section*{Author Contribution}

\sap\ is a large treasury program with the team work from many scientists, in particularly early-career researchers.
We include the following author contribution statements in compliance with \aastex\ \verb|v7.0| guidelines.
F.S.\ led the program design, observation planning, data analyses and writing of the paper. 
Y.F.\ led the imaging data reduction (Section~\ref{ss:3a_img}).
Y.F., X.L.\ and F.S.\ contributed to the photometric catalog (Section~\ref{ss:3b_cat}). 
F.S.\ led the grism data calibration and processing (Section~\ref{ss:3c_grism}).
X.L.\ led the photometric and spectroscopic redshifts analyses (Section~\ref{ss:4a_zph}, \ref{ss:4b_zsp}).
A.A., Y.F., J.M.H., T.H., X.J., X.L., F.S., W.L.T., Y.X.\ contributed to the visual inspection of spectra.
T.H.\ led the physical SED modeling (Section~\ref{ss:4c_sed}). 
J.M.H., X.L.\ and F.S.\ selected the $z>10$ galaxies candidates (Section~\ref{ss:5a_z10}).
F.S.\ led the ionized gas kinematics modeling  (Section~\ref{ss:5b_kin}).
X.L.\ discovered the dusty galaxy in Section~\ref{ss:5c_eml} and T.H.\ obtained the SED model.
Y.F., J.M.H., X.L., F.S.\ selected the AGNs in Section~\ref{ss:5c_eml}.
E.E.\ led the proposal writing, coordinated all efforts and oversaw the project as the PI. 
F.S., X.F., F.W.\ and J.Y.\ contributed to the proposal planning and writing.
X.F.\ named the survey.
All co-authors contributed to the scientific interpretation of the results and helped to write the manuscript.

% \begin{acknowledgments}

% \end{acknowledgments}

\vspace{5mm}
\facilities{JWST(NIRCam)}


\software{\textsc{astropy} \citep{2013A&A...558A..33A,2018AJ....156..123A}, 
\textsc{Bagpipes} \citep{carnall18},
\textsc{eazy} \citep{brammer08}, 
\textsc{emcee} \citep{emcee}, 
\textsc{FITSmap} \citep{hausen22}, 
\textsc{jwst} \citep{bushouse24},
\textsc{photutils} \citep{photutils},
\textsc{webbpsf} \citep{webbpsf}
}



\setcounter{figure}{0}
\renewcommand{\thefigure}{\thesection\arabic{figure}}

% \appendix

% \section{Lensing Magnification}
% \label{apd:01_lens}
% \input{90a_lens}


\bibliography{00_main}{}
\bibliographystyle{aasjournal}



% \suppressAffiliationsfalse
% \allauthors

\end{document}